\documentstyle[psfig]{l-aa}

\input epsf         
\begin{document}

\title{Star formation and evolution in accretion disks 
around massive black holes} 

\subtitle{Star formation and evolution in accretion disks}

\author{Suzy Collin\inst{1, 2}, Jean-Paul Zahn\inst{1}}

\institute{$^1$Observatoire de Paris, Section de Meudon, 92195 Meudon,
France\\$^2$Research Associate at Institut d'Astrophysique, Paris\\}

\date{Received ??? 1997; accepted ...} 

\thesaurus{02.01.2, 02.09.1, 11.01.2, 11.14.1, 11.17.3, 
12.05.1,19.37.1,19.92.1 
}

\offprints{Suzy Collin (Observatoire de Meudon)}

\maketitle{}
 

\begin{abstract}{We develop an exploratory model for the outer, 
gravitationally unstable regions of
accretion disks around massive black holes. We consider black 
holes  of mass 10$^6$ to 10$^{10}$ M$_{\odot}$, and primeval or 
solar 
abundances. In a first step we study star formation and evolution in 
a purely gaseous
marginally unstable disk, and we show that unstable 
fragments should collapse 
rapidly and give rise to compact objects (planets or protostars), 
which then accrete at a high 
rate and in less than 10$^6$ years acquire a mass of a few 
tens of M$_{\odot}$, according to a mechanism first proposed by Artymowicz, Lin and Wampler 
(1993). When these stars explode 
as  supernovae, the supernova shells break out of the disk, 
producing strong outflows. We show that the
gaseous disk is able to support a large number of massive 
stars and supernovae while staying relatively homogeneous. An interesting 
aspect is that the residual neutron stars can undergo other 
accretion phases, leading to other (presumably powerful) supernova 
explosions.
In a second step we assume that 
the regions at the periphery of 
the disk provide a quasi stationary mass inflow during the 
lifetime of quasars or of their progenitors, i.e. $\sim 10^8$ yrs, and 
that the whole
mass transport is ensured by the supernovae, 
which induce a transfer of angular
momentum towards the exterior, as shown by the numerical simulations 
of Rozyczka, Bodenheimer and Lin 
(1995). Assuming that
the star formation rate
is proportional to the growth rate of the  gravitational instability, 
we solve the disk structure and determine 
the gas and the stellar
densities, the heating being provided mainly by the stars 
themselves. We find self-consistent solutions in which the gas is 
maintained 
in a state very close  to  gravitational
instability,
in a ring located between 0.1 and 10 pc for a black 
hole mass of
$10^6$M$_\odot$, and between 1 and 100 pc for a black hole mass of 
$10^8$M$_\odot$ or larger, whatever
the abundances, and for relatively low accretion rates ($\le 10\%$ of the critical accretion
rate). For larger accretion rates the number of stars becomes so 
large that they
inhibit any further star formation, and/or the rate of 
supernovae is so high that they distroy the homogeneity and the marginal stability of the
disk. We postpone the study of this case.

Several consequences of this model can be envisioned,
 besides the fact that it proposes a solution to the 
problem of  the mass transport in the intermediate region of the 
disk where global instabilities do not work. 
As a first consequence, it could explain the high  
velocity metal enriched outflows implied by
the presence of the broad absorption  lines in quasars. As a second 
consequence 
it could account for a  pregalactic enrichment of the 
intergalactic medium, 
if black holes formed early in the 
Universe. Finally it could provide a triggering mechanism for starbursts in the central regions
of galaxies. A check of the model would be to detect a supernova 
exploding
within a few parsecs from the center of an AGN, an observation which 
can be performed in the near future.}


\end{abstract}

\section {Introduction.}

It is now a paradigm to state that quasars and Active Galactic Nuclei 
(AGN) harbour massive black holes in their centers (Rees 1984).
Massive black holes exist also at the center of
quiescent galaxies like ours. In quasars and AGN the accretion rate 
amounts
to a fraction of the  critical one. This rate of accretion of this 
black hole
must be sustained during at least 10$^8$ yrs to account for the total 
mass of the black holes locked in quasars and for the fraction of AGN 
among all 
galaxies (cf. for instance Cavaliere \& Padovani 1988 and 1989). Since the gas fueling the 
black hole at such a 
high rate cannot be produced 
inside the central parsec (except if there is a preexisting dense 
star cluster, through star
collisions, tidal disruptions, or ablation) a process for supplying 
gas from
larger distances is necessary, whatever it is. This will be a 
prerequisite of our
study.

There is a large concensus that AGN are fueled via 
accretion disks. Moreover the observation of the ``UV bump" (Shields 
1978, Malkan \& Sargent 1982, and many subsequent papers) argues in 
favor of geometrically thin and optically thick disks, possibly 
embedded in a hot X-ray emitting
corona. Generally these disks  are studied using the  $\alpha$ 
prescription for
viscosity introduced by Shakura and Sunayev (1973). It is 
well known that these ``$\alpha$-disks" have two serious problems at 
large radii: 
they are not able to 
transport rapidly enough the gas from regions located at say 
one parsec, and they
are gravitationally unstable
beyond about 0.1 parsec. They
should  therefore give rise to star formation, and as a consequence 
evolve rapidly towards stellar systems whose properties are quite 
different 
from gaseous accretion disks. 

The suggestion of star formation in the 
self-gravitating region of such an accretion 
disk has been first made by Kolykhalov 
and Sunayev
(1980). Begelman, Frank, and Shlosman (1989) and Shlosman and 
Begelman (1989)
discussed in more details the conditions for star formation at about 
1-10 pc from the black hole,
and its consequences on the disk. They concluded that unless 
the disk can be maintained in a hot or highly turbulent state, it 
should transform
rapidly into a flat stellar system  which will be unable to build  a 
new gaseous
accretion disk and to fuel a quasar (note that they apparently did 
not take into 
account the influence of self-gravity on the 
scale height, which would even strengthen this conclusion). It is why 
they finally adopt the picture of a disk made of  marginally unstable
randomly moving clouds, where the ``viscosity" of the disk is 
provided by cloud 
collisions. Here we adopt the opposite view that if a marginally unstable 
fragments 
begin to collapse owing to a local increase of density 
for instance, the collapse will continue until a protostar is formed, 
unless the collapse time is larger than the characteristic time for 
mass transport in the disk.

Farther from the center the supply of gas can be achieved by 
gravitational torques or by global non axisymmetric 
gravitational instabilities. However this does not solve the 
problem of the mass transport in the intermediate region where the 
disk is locally but not
globally self-gravitating, and one cannot  avoid appealing for a 
mechanism to transport angular 
momentum. This can be achieved by magnetic torques if large 
scale magnetic fields are anchored in the disk. We shall assume here
that neither turbulent nor large scale magnetic fields are important 
in accretion disks, and examine whether it is possible to find a 
solution to the problem of the existence of a quasi continuous flow 
with a high accretion rate in the simple framework of star formation 
and its feedback 
to the disk.

 As stated above, the outer regions of $\alpha$-disks
are strongly gravitationally unstable  (Hur\'e 1998). This is why in 
Collin and Hur\'e (1998, hereafter referred as
Paper 1) another prescription was adopted,  where the self-gravitating disk is  maintained  in a
state of  marginal instability, as initially proposed by  Paczy\'nski (1978). In
such a disk it is assumed that the transfer of angular momentum is 
not due to turbulent 
viscosity, but for
instance to  collisions between clumps and/or to gravitational 
dissipation, which prevent their
collapse. The properties of this disk were determined (its scale  
height $H$, its
midplane density and temperature, $\rho$ and $T$, as  functions of 
the 
radius $R$)  in the
framework of a vertically averaged stationary  model. We want here 
to add 
a few comments.

First it is not clear whether a marginally unstable disk can stay 
homogeneous, if the marginal state is maintained by a 
self-regulating mechanism requiring energy liberated by the 
interaction of turbulent 
clumps or by their gravitational collapse. In this case the disk 
would probably be made of discrete
interacting clouds. In order that the disk stays homogeneous, 
the mechanism providing the heating and the angular momentum 
transport 
should not perturb too much the
gaseous component. An important part of this paper is devoted to
verify that this condition is fulfilled in our model. 

Second we stress that the assumption of stationarity is not 
constraining, as the properties of
marginally unstable disks depend very little on the accretion 
rate (in particular the
density profile depends only on the central mass), and moreover the 
model is a local one (i.e. the different rings are independent).
Variations of  the accretion rate with radius, due for instance 
to 
infall of gas, should simply 
induce small variations of the
surface density and of the midplane temperature. The only assumption 
linked to stationarity in Paper 1 was to take the luminosity of the 
central 
source as given by the accretion rate in the gravitationally unstable 
region. But it was shown that this external radiation has 
not a strong influence on the radial structure.

It is also worth noting that in Paper 1 the disk was actually not 
supposed to be exactly marginally
unstable, since all results were parametrized with $\zeta$, the ratio 
of the
disk self-gravity  to the vertical component of the central 
attraction. The results are valid
provided that this quantity is not much larger than  unity (for 
marginal stability,
$\zeta=4.83$, cf. later). Again we stress that it is a local 
quantity, which can vary from place to 
place in the disk.

In Paper 1 
the heating of the disk was assumed due to dissipation of the 
gravitational 
energy of the accreting flow (for which we use the  word ``viscous 
heating", although it
might not be due to viscosity) and to the external radiative heating. 
For a disk 
made of stars and gas, other sources of heating can play a role, and 
one should take them into account. Note finally that
the mechanism for momentum and mass transport was not specified in 
Paper 1.

Here our approach will be as follows. First we assume the 
existence of a marginally unstable gaseous disk, and discuss the 
formation 
of stars, their evolution, and the 
feedback of the stars. We show that the influence of the stars stays 
local, so they 
do not destroy
the disk. The advantages of studying a pure gaseous disk is that the 
discussion can 
be carried out analytically, and gives a framework for studying 
the more complex case of a disk made of stars and gas. Second we
build a self-consistent model of a disk made of stars and gas, where 
we 
assume that stars themselves provide the transport of angular 
momentum sufficient to 
maintain the
accretion rate. The gas density in this model is constrained by the fact 
that stars can
form only if the disk is not
depleted in gas, i.e. if the gas is at least marginally unstable, and 
if 
the gas density is not too high, leading to a rapid 
destruction of the gas by star formation. It implies that the 
gaseous fraction of the disk 
is marginally unstable. On the other hand the number of stars is also 
regulated by the fact that too many stars would inhibite star 
formation by tidal effect, and would destroy the disk through 
supernova explosions.

In Sect. 2 we study the birth of stars inside the accretion disk, 
assuming 
that either it consists of primeval gas, or it has a solar 
abundance. We discuss the evolution and the ultimate
fate of these stars in Sect. 3. Sect. 4 is devoted to a discussion of 
the  feedback of the
stars, and Sect. 5 to the overall behaviour of the ``star-gas" 
disk. In Sect. 6 we envision different consequences of the model.

\section{Star formation in a marginally unstable disk}

One finding of Paper 1 was that a marginally unstable disk
is molecular for $R \ge$ 0.1 pc even if it is illuminated by a 
central UV-X source of
radiation. Another conclusion was that the structure of disks 
consisting of primeval gas is
different from that of disks having solar abundances. Primeval disks 
are 
optically thin (in the 
sense of Rosseland or Planck mean), and they are flaring. As a result 
they 
are
irradiated by the UV-X photons produced by the inner regions of the 
disk, and the external radiative flux exceeds the
gravitational flux beyond a critical radius of the order of 10$^4 
R_{\rm 
S}$ (Eq. 42 of
Paper 1), where $R_{\rm S}$ is the Schwarschild radius $2GM/c^2$. In 
the following we shall
therefore use in the zero metallicity case the self-consistent 
solutions found for the flaring
irradiated disk (Eqs. 5, 43, 44 and 45 of Paper 1). In the solar 
metallicity case, the disk
is optically thick and is not flaring; therefore there will be no external irradiation unless
the disk is warped or a fraction of the central radiation is backscattered. Since the
efficiency of backscattering and the variation of the warping angle with the radius  would have
to be chosen arbitrary, we prefer to use the solutions for a purely viscously 
heated disk (Eqs. 5, 6, 7, and 8 of Paper 1). This is justified  by 
the fact that generally the 
external irradiation flux
exceeds the gravitational flux only at large radii, of the order of 
a few tenths of a parsec, owing to the
large optical thickness of the disk. We will
distinguish these two cases at each step of the discussion, and to 
simplify, we call them the ``primeval" and 
the ``solar" case.

In Paper 1 it was also shown that simple analytical
expressions can be used to model the vertically averaged 
structure. We shall refer to these
expressions. The radius was expressed in 
10$^4R_{\rm S}$, as
it is a natural physical parameter for black holes, and corresponds 
roughly to the onset of 
gravitational instability. In the present 
paper, we prefer to 
express it in parsec, $R_{\rm pc}$. The black hole mass 
(called $M_6$) will be
expressed in 10$^6$M$_{\odot}$, which we take as a typical mass for  
small or
growing black holes, and we
shall often refer to the two cases $M_6=1$ and $M_6$=100 (a typical 
quasar or AGN black hole mass).

We recall that the 
expressions of Paper 1 are not valid beyond a few tenths of a parsec 
in the solar case for $M_6$ =
1, and beyond a few parsecs for $M_6$=100, as the temperature falls 
to very low values. On the
contrary the expressions for the primeval disk are always valid up to 
a 
few parsecs. As we shall see in Section 5, these limitations do not 
hold for the gas-star disk, owing to the important heating induced by 
the stars. 

 Since the disk is marginally unstable, gravitationally bound 
fragments  can form
with a mass  $M_{\rm frag}$ of the order of $4\rho H^3$ (Goldreich 
\&  
Linden-Bell 1965), where
$\rho$ is the midplane density and $H$ the height of the disk (equal 
to half its thickness). The 
conditions necessary for the
collapse of a fragment are that the time scale for star formation, 
$t_{\rm form}$, and 
the cooling time, $t_{\rm
cool}$,  be smaller than the characteristic mass transport time in 
the disk, $t_{\rm trans}$ (in an
$\alpha$-disk it is the viscous time). These times are 
approximated in replacing derivatives by ratios of finite 
quantities.  

We recall first that we are dealing with a non magnetized or a weakly
magnetized disk, and in particular we assume that the magnetic 
pressure is smaller than the 
thermal
pressure. Thus even if the time for ambipolar diffusion is large 
because the disk is not
neutral but weakly ionized, and the magnetic field is therefore tightly bound to 
the 
gas, it  is not able to support the fragment
against collapse, contrary to what is generally assumed in molecular 
clouds.
 So $t_{\rm form}$ corresponds to the maximum growth rate of the gravitation
instability, which is a function of the Toomre parameter $Q$ (Toomre 1964, Goldreich and
Linden-Bell 1965) defined  as:

\begin{equation}
Q = {\Omega c_{\rm s}\over \pi G\Sigma},
\label{eq-Toomre}
\end{equation}

where $\Omega$ is the keplerian angular velocity, $\Sigma$ is the 
surface density, and $c_{\rm
s}$ is the sound speed.
 Adopting the same
formulation as in Paper 1:

\begin{equation}
 \zeta = {4\pi G\rho \over \Omega^2},
\label{eq-zeta}
\end{equation}
 
 $Q$ can be written:

 \begin{equation}
Q={2\sqrt{\zeta+1}\over \zeta}.
\label{eq-Q}
\end{equation}

According to Wang \& Silk (1994) $t_{\rm form}$ is: 

\begin{equation}
t_{\rm form} = \Omega^{-1} {Q\over\sqrt{1-Q^2}}.
\label{eq-tform}
\end{equation}

$Q=1$ corresponds to marginal stability. Provided $Q$ is not too close to unity,
$t_{\rm form}$ is not much larger than the freefall time $t_{\rm ff}\sim 1/\Omega$:

\begin{equation}
t_{\rm ff}\sim 4.5\times 10^{11} M_6^{-1/2} R_{\rm 
pc}^{3/2}\ {\rm s}.
\label{eq-tffbis}
\end{equation}

 For $t_{\rm cool}$, one has to distinguish between
the optically thick solar case and the optically thin irradiation 
dominated primeval case. 

In the primeval case it is simply:

\begin{equation}
 t_{\rm cool}\sim {kT\over \Lambda(T)},
\label{eq-tcool-primeval}
\end{equation}

where $\Lambda(T)$ is the molecular hydrogen cooling function (Eq. 13 
of Paper 1). It writes: 

\begin{equation}
 t_{\rm cool}=1.8\times 10^7\zeta^{0.7} (f_{-1}f_{\rm E})^{-0.7}\ 
{\rm s},
\label{eq-tcool-primevalbis}
\end{equation}

where $f_{-1}$ is a correction factor of the order of unity (see 
Paper 1),
accounting for the flaring ($\sim$ 30$\%$), for the proportion of the 
bolometric 
luminosity used to heat the disk (another factor $\sim$ 20$\%$), and 
for a possible 
limb darkening effect of the UV-X source. $f_{\rm E}$ is
the Eddington ratio (the bolometric to Eddington luminosity ratio: 
like 
in Paper 1 we have deliberately made the assumption that the 
Eddington ratio is equal to the
ratio of the  accretion rate to the
critical accretion rate, for a mass energy conversion 
efficiency of 0.1; it means that we do not consider the possibility 
of a very low efficiency like in advection dominated disks, as these 
models do not seem to apply to high accretion rates).

For the solar case (i.e. gravitationally heated disks), $t_{\rm 
cool}$ is equal to: 

\begin{equation}
 t_{\rm cool}\sim{ 8\pi{R}^{3}\ H\over 3GM\dot{M}}\ { \rho kT\over 
\mu m_{\rm H}}
\label{eq-tcool-solar}
\end{equation}

which writes:

\begin{equation}
 t_{\rm cool}=1.8\times 10^7{\zeta^{10/7}\over
(1+\zeta)^{5/7}} f_{\rm E}^{-4/7} M_6^{-3/7}R_{\rm 
pc}^{-3/7}\kappa_{\rm R}^{3/7}\ {\rm s} 
\label{eq-tcool-solarbis}
\end{equation}

where $\kappa_{\rm R}$ is the mean Rosseland opacity.

The mass transport time is:

\begin{equation}
 t_{\rm trans}\sim {2\pi R^2\rho H\over \dot{M}}.
\label{eq-ttrans}
\end{equation}

In the primeval case it becomes:

\begin{equation}
 t_{\rm trans}\sim 3\times 10^{13}{\zeta^{0.85}\over 
(1+\zeta)^{1/2}} f_{-1}^{0.15}f_{\rm
E}^{-0.85}M_6^{-1/2} R_{\rm pc}^{1/2}\ {\rm s},
\label{eq-ttrans-primeval}
\end{equation}

and in the solar case:

\begin{equation}
 t_{\rm trans}\sim 1.7\times 10^{12}{\zeta^{8/7}\over 
(1+\zeta)^{4/7}} f_{\rm
E}^{-6/7}M_6^{-1/7} R_{\rm pc}^{-1/7}\kappa_{\rm R}^{1/7}\ {\rm s}.
\label{eq-ttrans-solar}
\end{equation}

\begin{figure*}
\psfig{figure=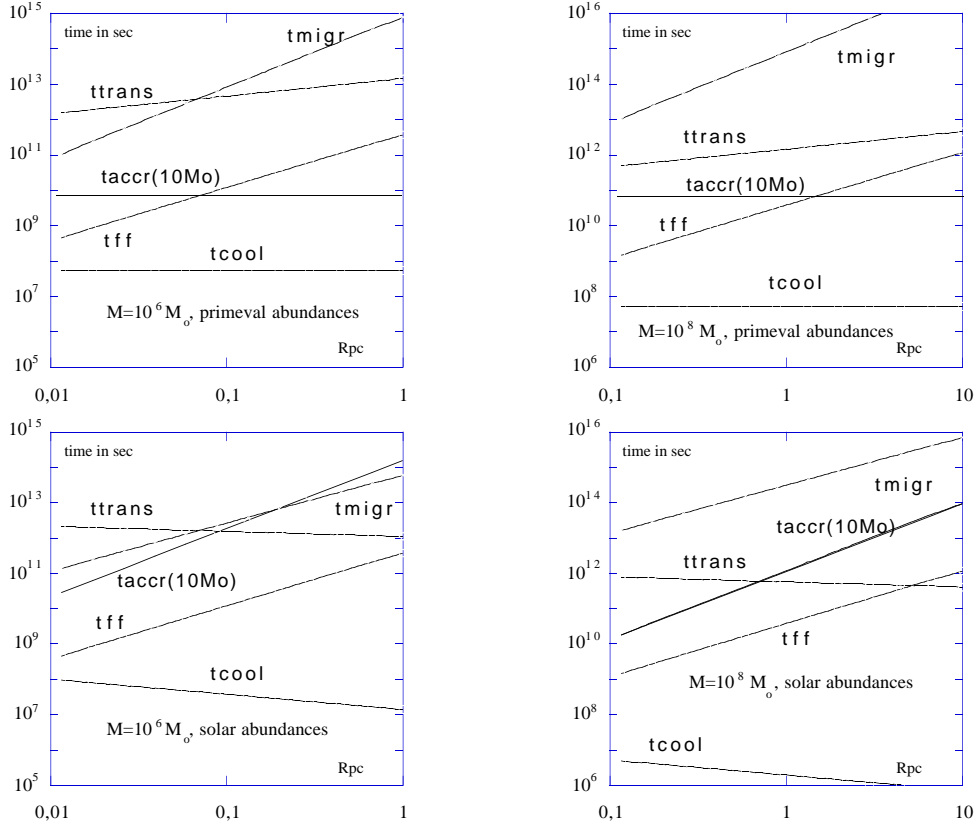}
\caption{different time scales in second, as functions of the radius 
in parsec, for
$M=10^6$M$_{\odot}$  and $M=10^8$M$_{\odot}$, and for a zero
and a solar metallicity; $f_{-1}$, $\kappa_{\rm R}$ and $f_{\rm E}$, 
have been set 
equal to unity, and $\zeta$ to 4.83. We recall that $t_{\rm acc}$ is 
actually overestimated.}     
\label{fig-time}
\end{figure*}

To make the discussion easier, these times are displayed on Figs. 
\ref{fig-time},  for
$M=10^6$M$_{\odot}$  and $M=10^8$M$_{\odot}$, and for a zero
and a solar metallicity. The factors
$f_{-1}$, $\kappa_{\rm R}$ and $f_{\rm E}$, have been set 
equal to unity, and $\zeta$ to 4.83, the value corresponding to the 
marginal instability $Q=1$.

Figs. \ref{fig-time} show that
$t_{\rm ff}$ and $t_{\rm cool}$ are smaller than $t_{\rm trans}$, so 
they satisfy the requirement for star
formation, up to a radius of the order of one parsec, where 
$t_{\rm ff}$ reaches $t_{\rm trans}$,
and therefore  the fragments  are drawn towards the black hole more 
rapidly than 
they collapse. Two results are worth to be noted. 
First $t_{\rm cool}$ is much
smaller than $t_{\rm ff}$ at the  beginning of the collapse (it
corresponds to the fact that the  ``equivalent" $\alpha$ is larger 
than unity). Thus a
strongly gravitationally unstable disk would 
immediately fragment into small
clumps. Second, $t_{\rm trans}$ is 
almost constant with
radius, in contrast with $\alpha$-disks, where it 
increases with radius and
becomes larger than the lifetime of the objects at a few tenths of a 
parsec. It 
means that the mechanism for
transporting angular momentum (which is not identified for the 
moment) must be very efficient at large 
radii. 

One should thus verify that
the radial drift velocity $V_{\rm rad}$ stays smaller than the 
Keplerian velocity $V_{\rm
Kep}$, or the disk would not be geometrically thin. 
The ratio $V_{\rm rad}/V_{\rm Kep}$ is equal to:

\begin{equation}
{V_{\rm rad}\over V_{\rm Kep}}\sim  (1+\zeta)^{1/2} {\dot{M}\over 
4\pi R^2 c_{\rm s}
\rho },
\label{eq-Vr/Vkep}
\end{equation}

which writes, in the primeval case:

\begin{equation}
{V_{\rm rad}\over V_{\rm Kep}}\sim 7\times 10^{-3} 
{(1+\zeta)^{1/2}\over \zeta^{0.85}} 
f_{-1}^{0.15} f_{\rm E}^{0.85} R_{\rm pc},
 \label{eq-Vr/Vkep-prim}
\end{equation}

and in the solar case:

\begin{equation}
{V_{\rm rad}\over V_{\rm Kep}}\sim 0.13  {(1+\zeta)^{9/14}\over 
\zeta^{8/7}}  f_{\rm
E}^{6/7}\kappa_{\rm R}^{-1/7} M_6^{-5/14} R_{\rm pc}^{23/14}.
 \label{eq-Vr/Vkep-solar}
\end{equation}

The ratio $V_{\rm rad}/ V_{\rm Kep}$ is therefore small up to large 
radii.

\medskip

For an isothermal and spherical collapse
the initial fragment would give rise to a dense core of mass $m_{\rm 
frag}$ in a time $t_{\rm collapse}$ corresponding to the growing rate
(Chandrasekhar 1939):

\begin{equation}
\dot{M}_{\rm collapse}\sim {c_{\rm s}^3\over G}.
\label{eq-rate-collapse}
\end{equation}

This rate writes, in the 
primeval case:

\begin{equation}
\dot{M}_{\rm collapse}=9\times 10^{-3}\zeta^{-0.45} (f_{-1} f_{\rm 
E})^{0.45}\ 
{\rm M}_{\odot}{\rm yr}^{-1}.
 \label{eq-rate-collapse-prim}
\end{equation}

and in the solar case:

\begin{eqnarray}
\dot{M}_{\rm collapse}& = & 1.5 \times 10^{-6} {\zeta^{3/7}\over 
(1+\zeta)^{3/14}} 
f_{\rm E}^{3/7}
\\
\nonumber
& \times  & M_6^{15/14} R_{\rm pc}^{-27/14} \kappa_{\rm R}^{3/7}\ 
{\rm 
M}_{\odot}{\rm yr}^{-1}
 \label{eq-rate-collapse-solar}
\end{eqnarray}

For the initial mass of the fragment, it corresponds to a time for 
the 
collapse of the core:

\begin{equation}
 t_{\rm collapse}\sim {\zeta\over(1+\zeta)^{3/2}}{1\over 4\pi\Omega}
\label{eq-time-collapse}
\end{equation}

which is
smaller than $1/\Omega$ for a marginally unstable gas.

Actually the collapse is not spherical, owing to the difference of disk
velocity between the
side facing the center and the opposite side, $\Delta V \sim 
H\Omega$. In a marginally unstable
disk, $\Delta V$ is of the order of the sound velocity, so the 
collapse begins quasi
spherically (this can be understood easily as the Bondi radius for 
spherical accretion by
an object of mass $m_{*}$,
$R_{\rm B} = Gm_{*}/c_{\rm s}^2$,  
and the Roche radius, 
$R_{\rm R} = R (m_{*}/M)^{1/3}$,  
are both equal to $H$ for the mass of the initial fragment).

In the equatorial plane the centrifugal force balances 
the gravitational
attraction. Therefore,
after having begun its collapse quasi spherically, the fragment will 
rapidly
lead to a condensed object (let us call it a protostar, although it 
does not have necessarily
the mass of a star, as we shall see below) and to a rotating 
protostellar disk.

Since the  small disk is detached from the rest of the disk, it 
cannot give its  angular
momentum to the external regions like in an ``infinite" accretion 
disk. So it must get rid
of a large fraction of its angular momentum, unless the final 
collapsed body would rotate with an
unacceptably large velocity. Indeed if the angular momentum is 
conserved  the angular velocity of
the protostar, $\Omega_{\rm final}$, is, according to the 
conservation of the angular momentum:

\begin{eqnarray}
\Omega_{\rm final} & \sim & \Omega_{\rm initial}
\left({\rho_{\rm final}\over \rho_{\rm
initial}}\right)^{2/3}\\
\nonumber
& \sim & 0.7 M_6^{-1/6} R_{\rm pc}^{1/2}\ \ {\rm s}^{-1}
\label{eq-angular-momentum}
\end{eqnarray}

where $\Omega_{\rm initial}$ is the average initial angular velocity  
of a fragment of radius
$H$, assumed equal to $\Omega(R)$, $\rho_{\rm initial}$ and $\rho_{\rm
final}$ are  the average initial and final densities, the latter 
being taken equal to 1 g
cm$^{-3}$. 

 The maximum angular velocity of a star, $\Omega_{\rm
max}$, is of the order of 5 $10^{-4}$ s$^{-1}$, for a density equal 
to 1 g
cm$^{-3}$. Actually the density is
higher for small mass stars, so $\Omega_{\rm max}$ is underestimated, 
while $\Omega_{\rm final}$
is overestimated as $\Omega_{\rm initial}$ is actually smaller than 
$\Omega$. Nevertheless there
are about 3 orders of magnitudes between $\Omega_{\rm final}$ and 
$\Omega_{\rm max}$.

A large proportion of the angular momentum is given to the 
protostellar disk, and in
fine also to a planetary system which is likely to form around the 
protostar. Since
$t_{\rm cool}$ is smaller than $1/\Omega$ at the  beginning of the 
collapse, it is indeed
probable that the initial cloud begins to fragment into smaller 
entities. 
Fragmentation stops when the cooling time 
(which increases during the 
collapse as the cloud becomes more optically thick), becomes of the 
same order as the collapse time. A fraction of the small fragments 
will form an ensemble of planetoids orbiting around the 
collapsed body, which can easily retain 99$\%$ of the angular 
momentum. 

Another way to evacuate the angular 
momentum of the small disk is to transform it into orbital motion in 
binaries or multiple
systems. One can also show that under the tidal action of the central 
mass the  small disk is synchronized in a time of the order of the 
time 
scale for dissipative friction,
i.e. in about a dynamical time. This mechanism can lead to the 
suppression of another large fraction
of the angular momentum. 

The actual final angular velocity, $\Omega_{\rm final\ real}$ is therefore 
at most of the order
of 10$^{-3}\Omega_{\rm final}$ given by Eq. 
\ref{eq-angular-momentum}. As the protostellar disk
gets rid of a large fraction of its angular momentum, the radius down 
which the collapse is
quasi spherical, $r_{\rm sph}$, is:

\begin{equation}
r_{\rm sph}  \sim \left({Gm_{\rm frag}\over \Omega_{\rm 
final\ real}^2}\right)^{1/3}\sim H {M\over m_{\rm frag}}\left({H\over 
R} \right)^3\sim 10^{-6} {H\over \zeta}
\label{eq-rsph}
\end{equation}

where we have assumed that the final angular momentum is of the order 
of 10$^{-3}$ times the
initial angular momentum. This equation shows that $r_{\rm sph}$ is 
of the order of the size of a
star, so the collapse proceeds as in the spherical case with a 
characteristic time $t_{\rm
collapse}$. 

\section{The accretion phase of the stars}

\subsection{Growth and evolution of the stars}

Once the protostar and the protostellar disk are formed, they undergo 
a mechanism of accretion
and growing proposed by Artymowicz, Lin and Wampler (1993).  These 
authors consider stars orbiting  
around  the central black hole, passing through the accretion disk, 
and finally being trapped
in the disk. They
showed that  these stars would accrete matter from the disk at a high 
rate 
and become rapidly massive stars which 
explode as supernovae
before being ``swallowed" by the black hole. They invoke this process 
to explain the metal rich
outflows observed in quasars (Broad Absorption Lines). 

There are several 
differences between Artymowicz et al. accretion process and ours. 
They too adopt 
the marginal instability
prescription,  but they do not solve the radial profile of the 
disk, 
taking the scale height as a given parameter. They do not consider 
the effect of replenishment of the gap or of the cavity opened by the 
stars in the
disk. Therefore the stars accrete only the gas which is initially 
inside the Bondi radius, and in a lesser extent inside the Roche 
radius. 

Note first that the same remarks as made previously for the collapse 
phase concerning the angular
momentum hold for the accretion phase, since the specific angular momentum of 
the 
matter located entering a volume $\sim H^{3}$ is of
the same order as that of the initial fragment. 

Due to the marginal instability prescription, the Roche radius 
$R_{\rm R}$  and the Bondi radius
$R_{\rm B}$ are equal to $H$ for
the initial mass of the star. After, $R_{\rm R}$
grows as $m_{*}^{1/3}$. So an underestimation
of the accretion rate can be obtained, assuming that it is limited to 
a region of radius $H$,
and is very roughly given by the shear velocity of this region:

\begin{equation}
\dot{M}_{\rm accr}\sim \Delta V \rho H^2 \sim \Omega m_{\rm frag}.
\label{eq-maccr}
\end{equation}

The corresponding time for a star to reach 10M$_{\odot}$,
$t_{\rm accr}(10{\rm  M}_{\odot})$, is equal equal to 
10M$_{\odot}/(\Omega m_{\rm
frag})$. One finds in the primordial case:

\begin{equation}
t_{\rm accr}\sim 10^{11}{(1+\zeta)^{3/2}\over 
\zeta^{0.55}} (f_{-1}f_{\rm
E})^{0.45}R_{\rm pc}^{-3}\ {\rm s},
\label{eq-taccr-prim}
\end{equation}

and in the solar case:

\begin{equation}
t_{\rm accr}\sim 6\times 10^{14}{(1+\zeta)^{2/7}\over 
\zeta^{10/7}} f_{\rm
E}^{-3/7}M_6^{-15/14} R_{\rm pc}^{-15/14}\kappa_{\rm R}^{-3/7}\ {\rm 
s}.
\label{eq-taccr-solar}
\end{equation}

We recall that these times are {\bf overestimations} of the real ones 
since we have
neglected the mass inflow towards the star coming from beyond $H$ and 
swept up by the Roche 
lobe, which unfortunately we cannot estimate simply. Since the Roche 
radius is commonly equal to 10 or 100 $H$ at the end of the accretion 
phase, we see that the accretion time can be overestimated by a very 
large factor. 

Fig.\ref{fig-time} displays $t_{\rm accr}(10{\rm M}_{\odot})$
as a  function of the radius. It is always small (10$^{2}$ to 
10$^{3}$ yrs) in the primordial case (owing to the large
value of the scale height). In the solar case it can reach a few 
10$^{6}$ yrs, i.e. be comparable to the evolution time of the star on 
the main sequence.

Palla \& Stahler (1993) have shown that stars with masses greater 
than 8 M$_{\odot}$ have
almost no pre-main sequence phase, as they are already burning 
hydrogen in the protostellar phase.
These stars reach the main sequence in typically 10$^4$ yrs if they 
are accreting at a rate
of 10$^{-5}$ M$_{\odot}$ yrs$^{-1}$. If 
accretion proceeds at a rate
of 10$^{-4}$ M$_{\odot}$ yr$^{-1}$, this time is increased as stars 
are not thermally
relaxed when the accretion is stopped, because they have larger 
radii. 

The study of Palla
\& Stahler is performed for solar abundances. One can estimate very 
roughly the effect of a
lower metal abundance. The effect of a
decrease of the abundances is similar to a decrease of the accretion 
rate. For a metallicity
10$^{-2}$ solar, one expects therefore that massive stars will not 
undergo a Kelvin-Helmoltz
phase even if the accretion rate is as large as 10$^{-3}$ M$_{\odot}$ 
yr$^{-1}$, and
that they will reach the main sequence in less than 10$^4$ yrs. For 
very high masses the
accretion rate is Eddington limited, and is at most of the order
of 10$^{-3}\times r_*/r_{\odot}$ M$_{\odot}$ yr$^{-1}$ (it is smaller 
if electron scattering does
not dominate on line absorption). Since this rate is larger than  
$\dot{M}_{\rm accr}$ this is actually not a limitation.

The growth of the stars is stopped when
they begin to evolve on the main sequence and to  drive strong 
radiatively accelerated winds.
These winds slow down and stop the accretion, as they transport a 
larger outflowing momentum flux
than the inflowing one. Outflowing winds are observed from Wolf-Rayet 
stars, up to 10$^{-4}$
M$_{\odot}$ yrs$^{-1}$, with velocities larger than 1000 km s$^{-1}$, 
but the efficiency of
radiative acceleration is probably smaller if the gas does not 
contain heavy elements
(Meynet et al, 1994).  Actually a rate of 
10$^{-6}$ M$_{\odot}$ yrs$^{-1}$
with such a velocity is amply  sufficient to stop the accretion.
 Moreover we will show in  the next section that it is not the wind, 
but the ionizing radiation
produced by the star, which stops the accretion, because it creates a 
cavity filled with hot
and dilute gas preventing further accretion.

\subsection{Gap opening and migration of the stars}

To obtain supernova explosions (which are required by our 
model, see Sect. 5), two
conditions should be  fulfilled: 1. the stars must grow up to at 
least 10 M$_{\odot}$, 
which implies that they 
must not open a gap in the disk,  
and 2. the stars must not be swallowed by the black hole before they 
reach the stage of supernova.
These conditions are both linked to the mechanism of tidal induced 
density waves discussed by Goldreich 
and Tremaine (1980), Ward (1986, 1988), and Lin and Papaloizou (1986a 
and 1986b).

Lin and Papaloizou showed that in an $\alpha$-disk a gap is 
opened by the 
star if its mass exceeds $m_{*{\rm gap}}$, where:

\begin{eqnarray}
 m_{*{\rm gap}} & \sim & M  {40\nu\over R^{2}\Omega} \left({H\over 
 R}\right)^{3/2}
 \\
 \nonumber
 &\sim & M {(40\alpha)^{1/2}\over 1+\zeta} \left({c_{\rm s}\over 
\Omega 
 R}\right)^{5/2}.
\label{eq-opengap}
\end{eqnarray} 

where $\nu$ is the kinematic viscosity. This expression is valid also 
in our 
case since the turbulent viscosity is still present and erases 
the gap. 
In the primeval case, it leads to: 

\begin{eqnarray}
 m_{*{\rm gap}} & \sim & 4 \times 10^3 
\alpha^{1/2}{\zeta^{-0.38}\over 
(1+\zeta)}
\\
\nonumber
& \times & 
(f_{-1} f_{\rm E})^{0.38} M_6^{-1/4}R_{\rm 
pc}^{5/4} \ {\rm M}_{\odot},
\label{eq-opengapprim}
\end{eqnarray} 

and in the solar case to:

\begin{eqnarray}
 m_{*{\rm gap}} & \sim & 2.6\ \alpha^{1/2} {\zeta^{5/14}\over 
(1+\zeta)^{33/28}}
\\
\nonumber
& \times & 
 f_{\rm E}^{5/14} \kappa_{\rm R}^{5/14} M_6^{9/14}R_{\rm 
pc}^{-5/14}\ {\rm M}_{\odot}. 
\label{eq-opengasolar}
\end{eqnarray}

 We see that in the primeval 
case  no gap is opened
for $M_{6}=1$ and $R \ge 0.01$ pc, and for $M_6=100$ and $R 
\ge 0.03$ pc. In the solar case a gap is always opened 
for $M_{6}=1$, while for $M_{6}=100$ no gap is opened at any radius. 
In other terms, star formation is not possible for small black 
hole masses and solar abundances in a marginally unstable purely gaseous 
disk. 
 
To know whether stars can reach the state of supernovae, one should
compare the evolution time scale to the migration time towards the 
black
hole due to the same
mechanism of induced  density waves. The migration proceeds towards
the interior if the midplane temperature decreases with increasing 
radius, which is our case.
The characteristic time for migration is: 

\begin{equation}
 t_{\rm migr} \sim  C\ {M^2\over m_{*}\Omega \pi R^2\Sigma}\ 
 \left({c_{\rm s}\over \Omega R}\right)^2
\sim  C\ {M\over m_{*}} {1\over \Omega}{H\over R},
\label{eq-migration}
\end{equation} 

where $C$ is a factor of the order of unity. 
Since $ t_{\rm migr}$ is inversely proportional to $m_*$, it means 
that the stars will begin to
migrate only when they will have acquired large masses. For $m_*$ = 
10 M$_{\odot}$, it gives,
in the  primeval case:

\begin{equation}
 t_{\rm migr}(10{\rm M}_{\odot}) \sim  2.4\times 
10^{15}{\zeta^{-0.15}\over (1+\zeta)^{1/2}}
(f_{\rm E}f_{-1})^{0.15} R_{\rm pc}^2\ {\rm s},   
\label{eq-migration10-prim}
\end{equation} 

and in the solar case:

\begin{eqnarray}
 t_{\rm migr}(10{\rm M}_{\odot}) & \sim  & 1.2\times 
10^{14}{\zeta^{1/7}\over (1+\zeta)^{4/7}}
\\
\nonumber
& \times & f_{\rm E}^{1/7} M_6^{5/14}R_{\rm pc}^{19/14}\kappa_{\rm 
R}^{1/7}\ {\rm s}.   
\label{eq-migration10-solar}
\end{eqnarray} 

Fig. \ref{fig-time} displays  $t_{\rm migr}(10{\rm M}_{\odot})$ as a 
function of the radius. It is of the
order of the evolution time of a 10 M$_{\odot}$ star ($\sim$ a few 
10$^{14}$ 
s, see Meynet et al. 1994) only for $R\ge 1$ pc.
This is a little too restrictive since star formation does not occur 
for $R \ge 1$ pc for a 10$^6$M$_{\odot}$ black hole.
However $t_{\rm migr}$ is larger  when a cavity is opened around the 
star, as it is the case here for massive stars, whose winds and HII 
regions excavate the disk 
during the major fraction of their life (see the next section). One 
can 
therefore consider 
that the migration time is not a constraint. 

\section{Feedback of stars and of supernovae on the disk}

To be complete the above picture of star formation and accretion should take into account the
effect  of massive
stars and of supernovae explosions on the disk. It is in particular 
necessary to
know the distance over which the influence of a massive star or a 
supernova
can be felt, and how long it lasts.
 
Many studies deal with star formation and
evolution in molecular clouds.  These problems should however be 
rediscussed in the framework of
accretion disks, since the conditions are not similar to molecular 
clouds (the density and the
temperature of the disks are higher, and they have strong Keplerian shear).  
Moreover Keplerian
disks are not isotropic: their scale height 
is much smaller than their
radius, so the disturbances can easily reach the surface in the 
``vertical" direction.
In the plane of the disk the 
Keplerian rotation interfers with an outflow when its velocity equals
the shear velocity, and it 
stretches the flow in the
azimuthal direction. 

Massive stars produce both ionizing photons which create HII regions, 
and winds, which create
pressure driven bubbles. As a result a cavity containing a 
dilute gas
expands around the stars. Shull (1980a) has shown that the bubble 
induced by the wind
should dominate over the compact HII regions during the first 10$^4$ 
years of the
life of the O stars. After this time, the expansion of the bubble is 
slowed
down, and it is overcome by the HII region. Thus the supernova 
explodes
in a low density cavity produced by the wind and the HII region. These
processes have generally been considered separetely, owing to the 
incertainties on the involved
phenomena (when in the O star life starts the wind and which is 
mass loss rate? Does it last
until the explosion of the supernova?...). Another problem lies in 
the fact that when the
radius of the bubble, the HII region, or the supernova remnant, is of 
the order of the scale
height of the disk, their evolution is strongly affected by the 
escape of material in the ``vertical"
direction. We shall study these effects using very crude 
approximations. 

\subsection{HII regions}

When an O star begins to shine on the main sequence, the initial 
radius of the HII region, $R_0$, is given by the equality
between the flux of ionizing photons and  the rate of recombinations:

\begin{equation}
R_0= \left({3\over 16\pi} {F_{\rm ion}\over \alpha_{\rm rec} 
n_0^2}\right)^{1/3}
\label{eq-R0}
\end{equation} 

where $\alpha_{\rm rec}$ is the recombination coefficient, $n_0$ is 
the number density in the
neutral medium (here molecular). Expliciting the variation of the 
density with the radius in the
disk, it writes:

\begin{equation}
R_0= 9.3\times 10^{15}\ 
\zeta^{-2/3}F_{49}^{1/3}T_4^{1/6}M_6^{-2/3}R_{\rm pc}^2\ {\rm cm},
\label{eq-R0bis}
\end{equation} 

where $F_{49}$ is the flux number of ionizing photons expressed in 
10$^{49}$ cm$^{-2}$ s$^{-1}$, and $T_4$ is the
temperature of the HII region in 10$^4$K. It is easy to see that this 
radius is always smaller
than the scale height of the disk in the region of interest for us.

The pressure in the HII region being larger than the pressure in the 
ambient medium, after a
dynamical time the Stromgren
sphere begins to expand at a rate (Spitzer 1978):

\begin{equation}
R_{\rm i}= R_0\ \left(1+{7\over 4} {c_{\rm s, HII}\ t\over 
R_0}\right)^{4/7}.
\label{eq-Ri}
\end{equation} 

Here  $c_{\rm s, HII}$ is the sound speed in the HII region. 
During this expansion the density
decreases as $R_{\rm i}^{-3/2}$, assuming that the ionizing photon 
flux stays constant. Since the
temperature of the HII region is constant, its pressure decreases, 
and when it reaches the
pressure of the ambient medium, for a radius $R_{\rm i,max}$, the 
expansion of the HII region is stopped. 

The ratio of the maximum radius $R_{\rm i,max}$ of the HII region to 
the scale height is thus:

\begin{eqnarray}
{R_{\rm i,max}\over H}& = & 0.46\ 
F_{49}^{1/3}T_4^{5/6}{(1+\zeta)^{1/2}\over \zeta^{0.364}}
\\
\nonumber
& \times & (f_{-1}f_{\rm
E})^{-0.505} M_6^{-1/6}R_{\rm pc}^{1/2} 
\label{eq-Rimax/H-prim}
\end{eqnarray}

in the primeval case, and:

\begin{eqnarray}
{R_{\rm i,max}\over H}& = & 200\ 
F_{49}^{1/3}T_4^{5/6}{(1+\zeta)^{2/3}\over \zeta^{13/21}}
\\
\nonumber
& \times & f_{\rm E}^{-1/4} \kappa_{\rm R}^{7/21}M_6^{-1/21}R_{\rm 
pc}^2 
\label{eq-Rimax/H-solar}
\end{eqnarray}

in the solar case.
It is of the order of unity in the primeval case, and can be much
larger in the  solar case
for $R \sim $ 0.1pc.

When the HII region reaches the surface of the disk  (i.e. when 
$R_{\rm 
i} \sim H$), the ionized  
material is allowed to expand above the disk in a low density medium. 
This is similar to the
``champagne" or ``blister" model of HII regions (see Tenorio-Tagle 
1979 and subsequent
papers), except that in the champagne model the HII region expands in 
the
intercloud  medium whose pressure is assumed equal to
that of the molecular cloud, while here it expands in a  
medium of very low pressure compared to the disk pressure. After a 
time $H/c_{\rm s}({\rm HII})$ much shorter than the life time $t_{\rm 
lt}$ 
of the star, the HII region is depressurized and maintained at the 
pressure 
of the ambient medium. {\bf The
ionization front stalls therefore at a radius $H$}. 
When it reaches the surface of the disk, matter is ejected from the
disk  with
a velocity  $c_{\rm s}({\rm HII})$. This matter is rapidly 
deccelerated by the 
gravity and returns to the disk (but not necessarily at the same 
place, since the ejection is not purely vertical) after about one 
rotation, and after having
travelled  on about $c_{\rm s}({\rm HII})/c_{\rm s}\times H \sim 
10H$. The result is therefore
very  different from the champagne
model, where the excavation process goes on after the HII region has 
opened a hole in the disk, and leads ultimately to the destruction of 
the whole molecular cloud.

In conclusion, the size of the HII region is equal to min($H, R_{\rm  
i,max}$) and is thus of the order of $H$.

\subsection{Winds}

O stars have strong winds with terminal velocities of the order of a 
few 10$^3$ km s$^{-1}$, through which they can eject up to 80$\%$ of 
their mass, 
if they reach the Wolf-Rayet stage. After a rapid phase of free 
expansion, the evolution is dominated by the ``snowplow" phase in 
which 
the ambiant gas is swept up, collapses into a thin shell and cools, 
and finally the flow settles into a steady state when the 
velocity of the shell equals the sound velocity of the ambient 
medium. 

The evolution of a wind driven 
``bubble"  has been studied by 
Castor, McCray and Weaver (1975) and subsequently by 
Shull (1980a). The major fraction of the life of the stars is spent in 
the snowplow 
phase, where a thin dense shell of radius $R_{\rm s}$ surrounds an 
inner region of stellar wind and of hot shocked wind. 
Shull has shown that after a time $t_{\rm cr}\sim  
10^{4} L_{36}^{-1/8} n_{5}^{-9/8} {\rm yr}$
 the pressure driving the shell falls owing to cooling, and the  
radius of the shell increases
as:
 
\begin{equation}
R_{\rm s}= R_{\rm cr}\times \left(1+{12\over 5}({t\over t_{\rm 
cr}}-1)\ + \ C_{1}({t\over t_{\rm cr}}-1)^2\right)^{1/4}
\label{eq-Rs}
\end{equation} 

where $R_{\rm cr}$ is the radius for $t=t_{\rm cr}$, equal to
$0.165 L_{36}^{1/8} n_{5}^{-7/8}$ pc, and $C_{1}$ is a nondimensional 
factor $\sim 2.7 \times L_{36} n_{5}^{1/4}$.
Here $n_5$ is the number density expressed in 10$^5$ cm$^{-3}$ and 
$L_{36}$ is the kinetic luminosity of the wind in 10$^{36}$ erg 
s$^{-1}$ (it corresponds to an outflow rate of $1.3\times 10^{-6}$ 
M$_{\odot}$ 
yr$^{-1}$ at a velocity of 2000 km s$^{-1}$).

 The expansion of the shell in the radial direction is stopped when 
its
velocity  is equal to the sound velocity or to the shear velocity, 
$\Omega\times 
R_{\rm s}$. 
The radius corresponding to the equality of the shear and of 
the expansion velocity, $R_{\rm  s,max1}$, is given by the solution 
of 
the equation:

\begin{equation}
{dR\over dt}(R=R_{\rm  s,max1}) = \Omega\times R_{\rm s,max1}.
\label{eq-Rsmax1}
\end{equation} 

Using the expression for the density in the disk as a function of
radius, one can easily show that only the linear term in Eq. 
\ref{eq-Rs} is important, and the solution can be written:

\begin{equation}
R_{\rm  s,max1}\sim 2 \times 10^{-2}\zeta^{-19/32}
L_{36}^{5/32}M_6^{-23/32}R_{\rm pc}^{69/32}\ {\rm pc}
\label{eq-Rsmax1bis}
\end{equation} 

The radius of the shell corresponding to an expansion velocity 
equal to the sound velocity, $R_{\rm  s,max2}$, is related to 
$R_{\rm  s,max1}$ by:

\begin{equation}
R_{\rm  s,max2}=R_{\rm  s,max1}\left({R_{\rm  s,max1}\over 
H}\right)^{1/3} 
(1+\zeta)^{-1/6}.
\label{eq-Rsmax2}
\end{equation}

It shows that when $R_{\rm  s,max1}$ is smaller than $H$, $R_{\rm  
s,max2}$ is smaller than $R_{\rm  s,max1}$, so the shell is stopped
at  the sound velocity  and not at the shear velocity, and vice versa.

In the primeval case, one has:

\begin{eqnarray}
{R_{\rm  s,max1}\over H} & = & 0.545 
{(1+\zeta)^{1/2}\over\zeta^{0.29}} 
(f_{-1}f_{\rm E})^{-0.303}
\\
\nonumber
& \times & L_{36}^{5/32}  M_6^{-7/32}R_{\rm pc}^{21/32},
\label{eq-Rsmax2-H-prim}
\end{eqnarray}

and in the solar case:

\begin{eqnarray}
{R_{\rm  s,max1}\over H} & = & 7.3 
{(1+\zeta)^{4/7}\over\zeta^{0.737}} 
f_{\rm E}^{-1/7}\kappa_{\rm R}^{-1/7}
\\
\nonumber
& \times & L_{36}^{5/32}  M_6^{-0.576}R_{\rm pc}^{1.30}.
\label{eq-Rsmax2-H-solar}
\end{eqnarray}

We see that $R_{\rm  s,max1}$ is larger than
$H$ only in the solar case and for small black hole masses, for which 
we have already found
that star formation is not possible. In the other cases  the  
expansion in the radial direction
is stopped at the sound velocity. Since it is the same in the 
azimuthal direction, the shell is
not streched and its expansion stays quasi isotropic  and 
 confined inside the disk. 

In conclusion
the influence of an O star does not extend much beyond a distance 
$H$, either through its HII
region or through its wind ``bubble". 

\subsection{Supernovae}

 The effect of a supernova is also to excavate the
disk and to create an expanding dense shell around the excavation.  
Several authors have studied the expansion of the shell 
inside a dense molecular cloud (Wheeler, Mazurek and Sivaramakrishnan 
1980, Shull 1980b, Terlevich et al. 1992). Here is added the 
complication of the orbital motion and of the small scale height of 
the 
disk.

As for the winds, the first expansion phase of the supernova 
is followed by a pressure modified snowplow phase which dominates the 
evolution. The 
expansion 
of the shell  proceeds then as (Shull 1980b): 

\begin{equation}
R_{\rm s} \sim 10^{-1} E_{51}^{1/4} n_{5}^{-1/2} \left({t\over t_{\rm 
sg}}\right)^{2/7}\ \ {\rm pc}
\label{eq-RsSN}
\end{equation} 

where $t_{\rm sg}$ is equal to 20$E_{51}^{1/8} n_{5}^{-3/4}$, and 
$E_{51}$ is the kinetic energy liberated by the explosion expressed 
in 
10$^{51}$ ergs.

Again the expansion in the radial direction is stopped when the 
expansion velocity is equal either to the sound velocity or to the 
shear 
velocity at its edge. For the shear velocity, it gives:

\begin{equation}
R_{\rm  s,max1}\sim 0.17 \zeta^{-2/7}
E_{51}^{3/14}M_6^{-3/7}R_{\rm pc}^{19/14}\ {\rm pc}.
\label{eq-Rsmax1SN}
\end{equation} 

One checks that $R_{\rm  s,max1}$ is larger than the scale height of 
the 
disk, both for the solar and for the primeval cases. As a 
consequence the 
radius corresponding to a velocity of the shell equal to the sound 
velocity is larger than $R_{\rm  s,max1}$ so the sound velocity does 
not play any role in 
stopping the shell in the radial direction. The shell can 
still expand in the azimuthal direction and will therefore be 
streched and sheared by the 
differential rotation. It will disappear after about one rotation 
time.

However since the shell expands always beyond a radius $H$ (Eq. 
\ref{eq-Rsmax1SN}), it breaks
rapidly  out of  the disk,  the 
interior is 
depressurized and its internal energy becomes negligible, so Eq. 
\ref{eq-Rsmax1SN} does not apply.  Then $R_{\rm s,max1}$ must be 
estimated assuming  that the
expansion is driven only by momentum conservation. A large fraction 
of that momentum escapes
from the disk, and the  momentum supplied to the disk by the 
supernova is finally:

\begin{equation}
P_{\rm disk}\sim P_{\rm tot} {H\over R_{\rm  s,max1}}
\label{eq-Pdisk}
\end{equation} 

where $P_{\rm tot}$ is the total momentum carried by the supernova 
explosion. Since the terminal velocity is of the order of $3P_{\rm 
disk}/(4\pi\rho H R_{\rm  s,max1}^{2})$, one gets:

\begin{equation}
R_{\rm  s,max1} \sim {3P_{\rm tot}\over 4\pi\rho\Omega}.
\label{eq-Rsmax1SNbis}
\end{equation} 

It gives:

\begin{equation}
R_{\rm  s,max1}=0.22 \zeta^{-1/4}
P_{43}^{1/4}M_6^{-3/8}R_{\rm pc}^{9/8}\ {\rm pc}
\label{eq-Rsmax1SNter}
\end{equation} 

where $P_{43}$ is expressed in 10$^{43}$ g cm s$^{-1}$. This is 
actually not very different from Eq. \ref{eq-Rsmax1SN}. 

 A numerical 2-D simulation of the explosion of a supernova in a thin
keplerian disk has been performed by  Rozyczka, Bodenheimer and Lin 
(1995). 
It shows how the shell is deformed and split into
two elongated edges, one corresponding to the leading hemisphere, 
the other
corresponding to the trailing hemisphere, the radial separation 
between the two 
edges being of the
order of $t\ \Delta V(R_{\rm  s,max1})$. There is however an important 
difference with 
us. Rozyczka et al. computation  does not depend on the scale height, 
but only on the surface
density which is given a priori. So they assume
that the disk receives the momentum corresponding to  the velocity 
vectors inclined by less than
$\pi/4$ to the mid-plane,  which means $P_{\rm disk}= P_{\rm 
tot}\times \pi/8$. This is generally
larger than our value for $P_{\rm disk}$ (as $H/R_{\rm  s,max1}$ is  
smaller than 
unity in our marginally unstable model). Consequently the angular 
momentum supplied by one 
supernova to the disk is smaller in our case.

\bigskip

In conclusion of this section we can estimate the maximum number of 
stars  that the disk can support at the same 
time without being too much perturbed: it is given by the maximum 
number of HII 
regions and wind bubbles which can fit into the disk.  
Let us define the number of stars per decade of radius as $N_{*}$. 
Since the radius of the HII regions and of the wind bubbles is of 
the order of or smaller than $H$, an {\bf underestimation} of the 
maximum number of stars is then $N_{\rm *max} $. 

\begin{equation}
N_{\rm *,max} \sim 2 \pi (R/H)^2
\label{eq-N*max}
\end{equation}

In the primeval case, it gives: 

\begin{equation}
N_{\rm *,max} \sim 2\times 10^3 (f_{1}f_{\rm 
E})^{-0.3}(1+\zeta)\zeta^{0.3} M_{6}R_{\rm pc}^{-1},
\label{eq-N*max-prim}
\end{equation} 

and in the solar case:

\begin{equation}
N_{\rm *,max} \sim 2\times 4\times 10^5 f_{\rm E}^{-2/7}\kappa_{\rm 
R}^{-2/7}{(1+\zeta)^{8/7}\over \zeta^{2/7}} M_{6}^{2/7}R_{\rm 
pc}^{2/7}.
\label{eq-N*max-solar}
\end{equation} 

We can also determine the maximum rate of supernovae per decade of 
radius supported by the
disk. As the cavities created by the blast waves are replenished 
roughly 
at the shear velocity, their lifetime is $\sim 1/\Omega$, and this rate 
is 
of the order of
$\Omega (R^2/R_{\rm  s,max1}^2)$, or:

\begin{equation}
{\cal{N}}_{\rm SN,max} \sim 1.4\times 10^{-3} P_{43}^{-1/2} \zeta^{1/2} 
M_{6}^{5/5}R_{\rm 
pc}^{-2}\ {\rm yr}^{-1}.
\label{eq-SNrate}
\end{equation} 

One could think that this number is overestimated because it does not 
take into account 
the stretching of the cavity in the azimuthal direction. One the 
contrary this estimation seems
rather conservative, as the numerical 
simulations of Rozyczka et al (1995) show 
that when the cavity and the shock wave reach their maximum radial 
extension after a fraction of an orbital time and become strongly 
elongated, they are also strongly 
squashed, so the surface of the perturbation seems to be rapidly 
decreasing.

The corresponding maximum number of stars, assuming they evolve in 
3$\times 10^6$ yrs, is
generally smaller than that given by Eqs. \ref{eq-N*max-prim} and 
\ref{eq-N*max-solar}. So
finally in most cases the supernovae, and not the HII regions and the 
wind bubbles, will limit
the number of stars that the disk is able to sustain.

\section{A model for a self-regulated  disk made of stars and gas}

We would like now to find whether it is possible for the stars and 
the gas to coexist in the
disk in a quasi stationary state. To perform this study we make 
several assumptions. The
first one, which has been discussed in the previous sections, is that 
when a fragment begins to
collapse, it leads   necessarily to the  formation of a massive star. A 
second basic assumption is that
the gaseous fraction of the disk stays quasi homogeneous. This is of 
course a very rough
approximation  considering the presence of HII cavities (although we 
will check that they occupy
a very small volume with respect to the whole disk in the models). But the 
most important problem in
this respect is to know what
kind of perturbation supernovae induce in the disk in the long term, 
after their blast
wave phase which is short owing to rapid erosion by the shear. It is
clear that only numerical 3-D simulations would allow to answer this 
question.

Another assumption is that the regions at the periphery of  
the disk provide a quasi stationary mass inflow
during the  life time of quasars or of their progenitors
(for instance via global gravitational instabilities induced by 
merging), 
equal to the accretion rate on the black hole. In other words we 
assume 
that there is neither infall on the inner regions of the disk, nor a
strong outflowing mass rate from the disk (except that due to the 
supernovae, which is small
with respect to the accretion rate in the following models). It would 
be actually possible to
relax these assumptions, but we postpone such a study.

\subsection{Angular momentum and mass transport}

 Another
difference with  the galactic disk is that we have to find an 
efficient mass transport 
mechanism, because the disk is an accretion one. Supernovae produce a 
net transfer of angular momentum 
towards the exterior, as discussed in the previous section and shown 
by the numerical
simulations of Rozycska et al. (1995). This is because the leading 
hemisphere of the supernova
shell has an excess of angular momentum compared to the disk, while  
the trailing  hemisphere has
a deficit of angular momentum. Consequently the  leading 
hemisphere is driven towards the center and the trailing one towards 
the exterior, as discussed already in the previous section. The net
angular momentum supplied by one
supernova is equal  to: 

\begin{equation}
\Delta J \sim {3\over 2\pi} P_{\rm tot} R {H\over R_{\rm  s,max1}}. 
\label{eq-angular-momentum-bis}
\end{equation} 

Rozyczka et al. (1995) estimate the angular momentum redistributed in 
the 
midplane 
of the disk by one supernova to be of the order of $(1/8)
P_{\rm tot}  R$ to the disk. We have seen that this estimation must 
be corrected in our
case by  a factor $\sim 4H/ R_{\rm  s,max1}$. Numerically Eq. 
\ref{eq-angular-momentum-bis}
gives:

\begin{equation}
\Delta J \sim 2\times 10^{43} P_{43}^{3/4} \zeta_{\rm g}^{1/4} 
M_{6}^{3/8}R_{\rm 
pc}^{-1/8} H_{\rm cm}\ {\rm g\ cm^{2}\ s^{-1}}.
\label{eq-SNrate}
\end{equation} 

(we distinguish from here on the contribution of the gas, $\zeta_{\rm 
g}$, and of the stars, 
$\zeta_{\rm *}$, to $\zeta_{\rm tot}=\zeta_{\rm g}+\zeta_{\rm *}$).

 In the absence of another
known mechanism it is tempting to identify the mechanism of momentum 
transport with that
induced by supernovae explosions. Let us assume therefore that {\bf 
all} the angular momentum
required to sustain the  accretion rate is carried by these 
explosions. The rate of supernovae
per decade of radius is then given  by the relation:

\begin{eqnarray}
 {\cal N}_{\rm SN} & \sim & {2 \dot{M} R^{2} \Omega\over \Delta J 
log_{10}e}
 \\
 \nonumber
 & \sim & {2.3\times 10^{14}\over H_{\rm cm}}\zeta_{\rm g}^{-1/4} 
P_{43}^{-3/4}  f_{\rm E}
M_{6}^{9/8} R_{\rm 
 pc}^{5/8}\ {\rm yr}^{-1}.
\label{eq-NSN}
\end{eqnarray} 

\subsection{Regulation of star formation}

 Now we have to look for a mechanism regulating the star
formation. Such mechanisms have been studied by several authors in  
the context of the Galaxy (McCray \& Kafatos 1987, MCKee 1989, Palous, 
Tenorio-Tagle \& Franco 1994 among others). The idea which prevails generally is that the stars 
themselves provide a physical
mechanism for induced star formation and regulation, via the  
creation of dense supernovae
shells and of HII regions inhibiting the formation of new stars and 
even destroying the clouds.
In our case supernovae may indeed induce star formation in the disk, 
as they lead to
overdensities which should trigger gravitational instabilities.  
Concerning the inhibition of
star formation, we will see that it is not due to the presence of HII 
regions which stay limited
in size, but to the mass of the other stars.

A self-regulation mechanism for the gas density has been proposed for 
the 
Galactic disk by Wang and 
Silk (1994), through the growth time of gravitational instabilities.
We adopt their view that the rate of gas transformed into stars
is:

\begin{equation}
{d\Sigma_{\rm g}\over dt}= \Sigma_{\rm g}{\epsilon \over \Omega} 
{\sqrt{1-Q^2}\over Q},
\label{eq-taux-star-form}
\end{equation} 

 $\epsilon$ being an ``efficiency'' of star formation which 
 ``parametrizes our ignorance". In the Galaxy 
it is of the order of  0.1$\%$ for massive stars.
 
Wang and Silk have included the stellar contribution in the Toomre 
parameter $Q$ of Eq. \ref{eq-taux-star-form}, as given by 
a two fluid approximation. Here it 
should not be taken into account,
since the mean distance between stars is larger than the instability 
length 
$H$. Indeed we recall that, owing to the HII and the wind bubble 
cavity, a new star can form only
at a distance greater than $H$ from an existing star.

An immediate consequence of Eq. \ref{eq-taux-star-form} is that the 
gas density should be close to the value corresponding to marginal 
instability, unless all the gas 
would be rapidly transformed into stars.
Another consequence is that the surface 
density of the stars should not be too large, unless star formation 
would be
inhibited by the tidal effect of nearby stars. This condition writes:

\begin{equation}
{\Sigma_{\rm g}\over \Sigma_{\rm *}}\ge {H\over d},
\label{eq-tidal-*}
\end{equation} 

where $\Sigma_{\rm g}$ and $\Sigma_{\rm *}$ are the surface densities 
respectively of gas and of stars and $d$ is the mean distance between stars. A way of expressing
this  condition is to say that $(\Sigma_{\rm g}^2/
\Sigma_{\rm *}^2)$ should be larger than  the ratio of the number of 
stars, $N_{*}$, 
to the maximum allowed number $N_{\rm *,max}$ given by Eq. 
\ref{eq-N*max}.

\subsection{The equations}

Since the formation time and the
lifetime of the stars are both small compared to the growth time of 
the black hole or to the active phase of a quasar,
a stationary  state of the disk could be established if 
there is a 
steady mass inflow from outward.  We can determine then the 
star formation rate from the number of supernovae able 
to sustain the required accretion rate, and we can solve the 
radial disk structure in a self-consistent way.

The (vertically averaged) equations for the disk structure are the 
the hydrostatic equilibrium equation:

\begin{equation}
c_{\rm s} = \Omega H (1+\zeta_*+\zeta_{\rm g})^{1/2},
\label{eq-hydro}
\end{equation} 

where we have included the stellar contribution, and the energy equation which we shall now
establish.

There are 3 heating mechanisms for the disk:

\noindent - Heating by the
fraction of the stellar luminosity not absorbed in the HII regions,
i.e. the near UV and visible spectrum  (actually a major fraction 
of the bolometric
luminosity of the stars). We shall make the rough approximation of an 
homogeneous heating, while
it is clearly not the case, since the mean distance between the stars 
is larger than the scale
height of the disk. This approximation overestimates the proportion 
of the stellar flux
absorbed in the disk, as it does not take into account geometrical 
factors. 

\noindent - Heating by the central source, when the disk flares. We shall see that it 
is now always the
case, even for solar abundances, contrary to the purely gaseous disk.

\noindent - Heating due to the dissipation of kinetic energy through 
shock waves 
induced by supernova shells (it corresponds to the viscous heating in 
a standard disk). 

The problem is to determine the proportion of 
photons absorbed in the disk. One of the difficulties comes from the 
fact that only average -
grey - opacities are generally considered in this kind of problem, 
in particular 
in Paper 1. Actually it would be necessary to know the relation 
between the opacity in
the far infrared, which gives the flux radiated by the disk, to the 
opacity in the optical-UV
range, which gives the fraction of the radiative flux absorbed by the 
disk. It is
out of the scope of this paper to perform detailed frequency 
dependent opacity computations,
but clearly the model would require such a study to be fully validated. Here we shall make
different assumptions in the solar and in the primeval case.

The energy equation is:

\begin{equation}
\sigma T^{4}=\sigma T_{\rm irr}^4+\sigma T_{\rm grav}^4.
\label{eq-heat}
\end{equation}

Here $\sigma T_{\rm irr}^4 = F_{\rm 
ext}\ +\ F_{*}$, where $F_{\rm ext}$ is the
 external flux ($F_{\rm inc}$ in Paper 1, given by Eq. 32), and 
$F_{*}$ the flux due
to the stars. $T_{\rm grav}$ is the usual contribution of the 
gravitational release.

For the primeval
case - we recall that the disk is then optically thin - we adopt the 
grey prescription as in
Paper 1. The energy equation writes thus:

\begin{equation}
\sigma T^4=  F_{\rm ext} + {4 N_* L_*\over\pi R^2} + 
{3\over2\pi}{\Omega^2\dot{M}\over\tau}, 
\label{eq-heat-prim}
\end{equation}

where $\tau$ is the optical thickness
corresponding to $H$,  $N_*$ is the
number of stars per decade of radius and $L_*$ is their average 
optical-UV luminosity.

For the solar case, we use a ``modified" grey approximation. In the 
first part
of the present paper where we have assumed a purely gaseous disk, we had only to deal with the
optically thick  region of the disk. This is no more the case, because we would like 
to extend the model
at larger radii, since stars maintain a temperature higher than in 
the pure gaseous case.
The disk becomes then optically thin in the sense of the Planck mean, 
but the major
part of the visible and UV luminosity  of the stars is however 
absorbed by dust
since the surface density of the disk is always larger than 10$^{21}$ 
cm$^{-2}$.
So finally the energy equation is, in the optically thick case:

\begin{equation}
\sigma T^4=  F_{\rm ext} + {4 N_* L_*\over\pi R^2}\tau + {9\over 
64\pi}\Omega^2\dot{M}\tau,
\label{eq-heat-solar-thick}
\end{equation}

and in the optically thin case:

\begin{equation}
\sigma T^4=  F_{\rm ext} + {4 N_* L_*\over\pi R^2\tau} + {2\over 
\pi\tau}\Omega^2\dot{M}.
\label{eq-heat-solar-thin}
\end{equation}

Note that $N_*$ can be expressed in terms of the supernova rate, as
 $N_*\sim t_{\rm lifetime}\cal{N}_{\rm SN}$, where $t_{\rm lifetime}$ 
 is the mean lifetime of the stars.

We further specify the rate of star formation per decade of radius (which 
is equal to the supernova rate
in a stationary disk):

\begin{equation}
{\cal{N}}_{\rm SN}={1\over m_{\rm frag}} \pi^{1/2} \rho_{\rm g} H R^2 
\epsilon\ {(1-Q^2)^{1/2}\over Q}
\Omega \ \zeta_{\rm g}^{1/2}
 \label{eq-star-form}
\end{equation}

where $ m_{\rm frag}$ is the mass of the most unstable fragments, 
$\sim \rho H^3$.
 $\epsilon$ must take into account 
 the gas returning to the disk through winds and
escaping from the disk through supernovae, as well as the mass of 
neutron stars which
migrate into the black hole (or are 
eventually ejected from the disk if the explosion is not 
symmetrical). We recall that {\bf $Q$ takes into
account only the gas density}, contrary to the galactic case. Note 
also that the actual 
parameter of the problem is not $Q$, but $\epsilon\sqrt{1-Q^2}/ Q$. 

Finally the star mass density is given by:

\begin{equation}
\rho_*\sim {\Sigma_*\over 2H} \sim { m_* t_{\rm lifetime}log_{10}e\ 
N_{\rm SN}\over 4 \pi H R^2}
 \label{eq-star-dens}
\end{equation}

where we assume that the stars and the gas have the same scale 
height.

The complete system consists of Eqs. \ref{eq-taux-star-form}, 
\ref{eq-hydro}, and \ref{eq-heat} (in the form of Eqs. 
\ref{eq-heat-prim}, \ref{eq-heat-solar-thick}, or 
\ref{eq-heat-solar-thin}), \ref{eq-star-form} and 
\ref{eq-star-dens}, plus the equations defining $\zeta_{\rm g}$ and 
$\zeta_*$ (Eq. \ref{eq-zeta}).

\subsection{The results}

We recall 
first that, due to our rough approximation of the opacity 
 coefficient, our solutions are valid only for a molecular disk, 
where the
cooling rate (in the primeval case) or the opacity coefficient 
$\kappa$ (in the solar case)
are close respectively to $\Lambda(T)$ and to 1 g$^{-1}$ cm$^{2}$. It corresponds 
to the range of temperature $T\le 1500$ K.
We have also
seen that our model is limited by several constraining hypothesis, 
such 
as the constancy in space and in time of the 
accretion rate. We will also limit the 
solutions to a marginally unstable gas. This will impose a 
condition on the efficiency factor $\epsilon$. It is however possible 
that other solutions, implying for instance a larger accretion rate, 
exist for an unstable gas, but we do not consider this case 
here. Our aim is actually only to show 
that even in this restricted framework, one can find solutions
fulfilling the conditions discussed in the previous sections.

\begin{figure*}
\psfig{figure=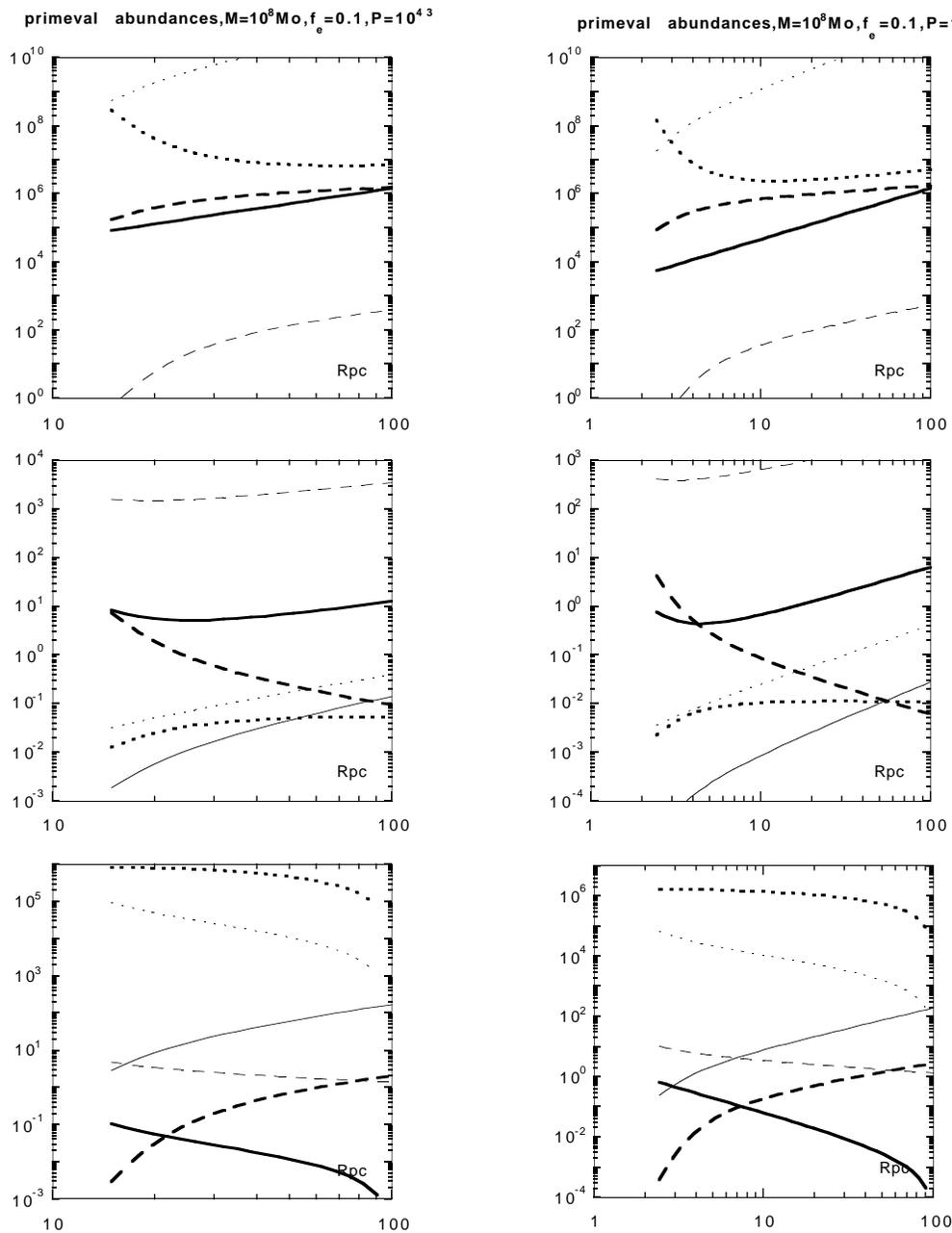}
\caption{Solutions for the disk made of stars and gas. In the top 
panels are given the characteristic times in year: bold solid 
lines: orbital 
time; bold dashed lines: $t_{\rm trans}$; bold dot lines: $t_{\rm 
accr}$(10M$_{\odot}$); thin dashed lines: $t_{\rm cool}$; thin dot lines: $t_{\rm 
migr}$(10M$_{\odot}$). In the middle panels, are given several parameters
allowing to check that the conditions required by the model are 
fulfilled: bold solid lines: ratio of the rate of supernovae to the 
maximum allowed rate; bold dashed 
lines: ratio  of the number of stars to the 
maximum number allowed by the tidal effect; bold dot lines: ratio  of 
the number of stars to the 
maximum number allowed by the HII regions; thin solid lines: 
$\epsilon\sqrt{1-Q_{\rm g}}$; thin dashed line: $m_{\rm gap}$ in 
M$_{\odot}$; thin dot lines: $V_{\rm rad}/V_{\rm Kep}$. In the bottom 
panels, several physical parameters of the models: bold solid 
lines: ${\cal{N}}_{\rm SN}$ per year, integrated from the outer edge; 
bold 
dashed lines: $m_{\rm frag}$ in M$_{\odot}$; bold dot 
lines: $M_{\rm gas}$ integrated from the outer edge; thin solid 
lines: $H/10^{16}$cm; thin dashed lines: $T/100K$; thin dot lines: 
$N_{*}$ integrated from the outer edge.}
\label{fig2-gaz-etoiles}
\end{figure*}

\begin{figure*}
\psfig{figure=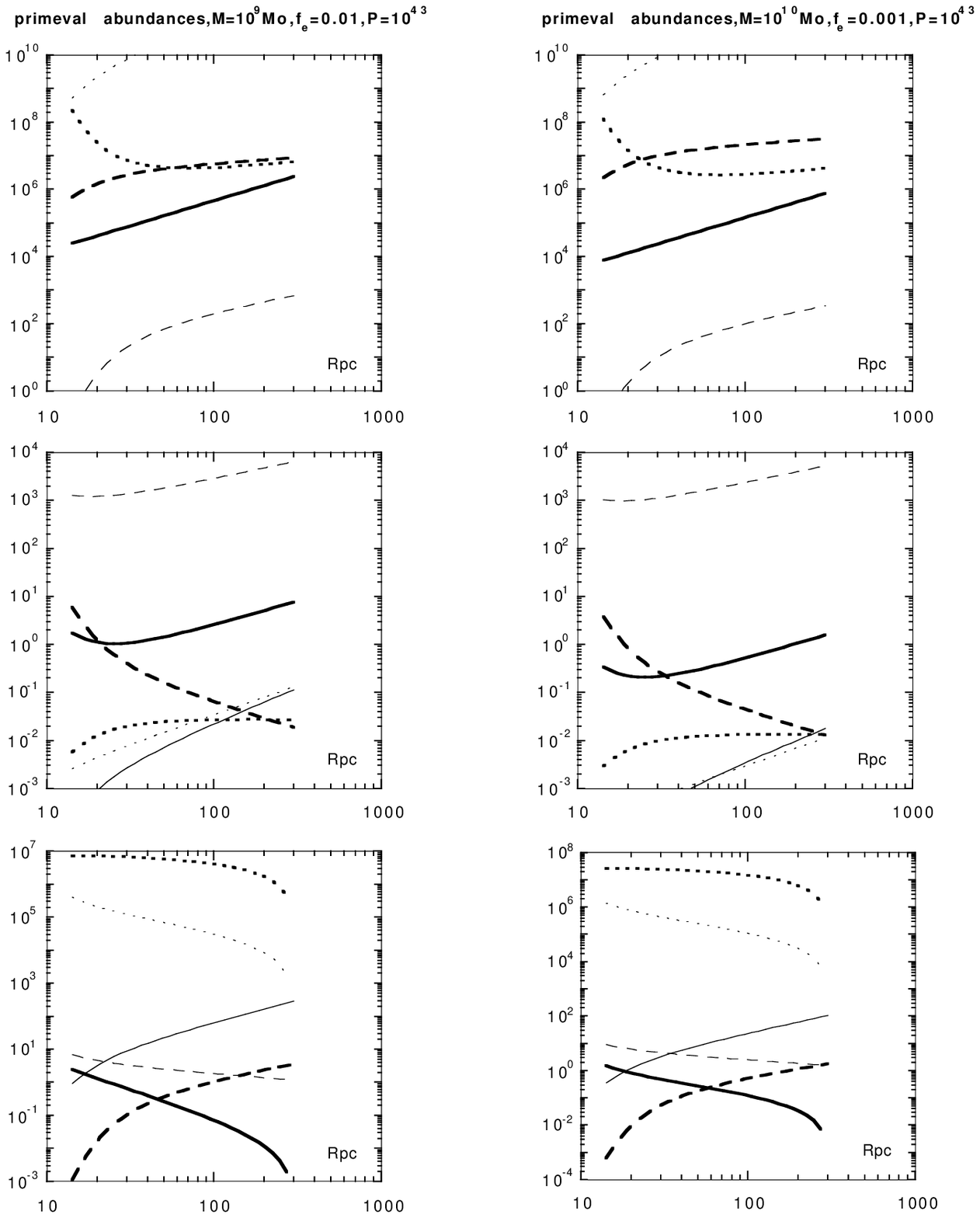}
\caption{Same as Fig.2, for other cases}
\label{fig3-gaz-etoiles}
\end{figure*}

\begin{figure}
\psfig{figure=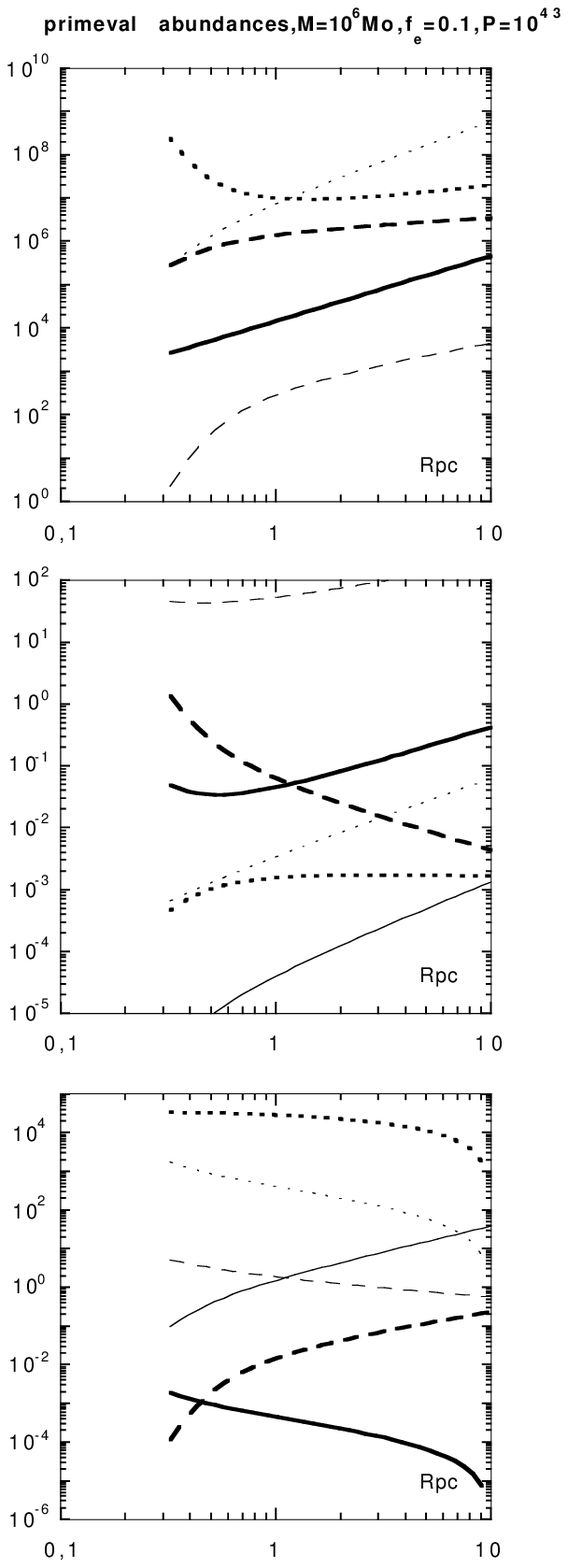}
\caption{Same as Fig.2, for other cases}
\label{fig4-gaz-etoiles}
\end{figure}

\begin{figure*}
\psfig{figure=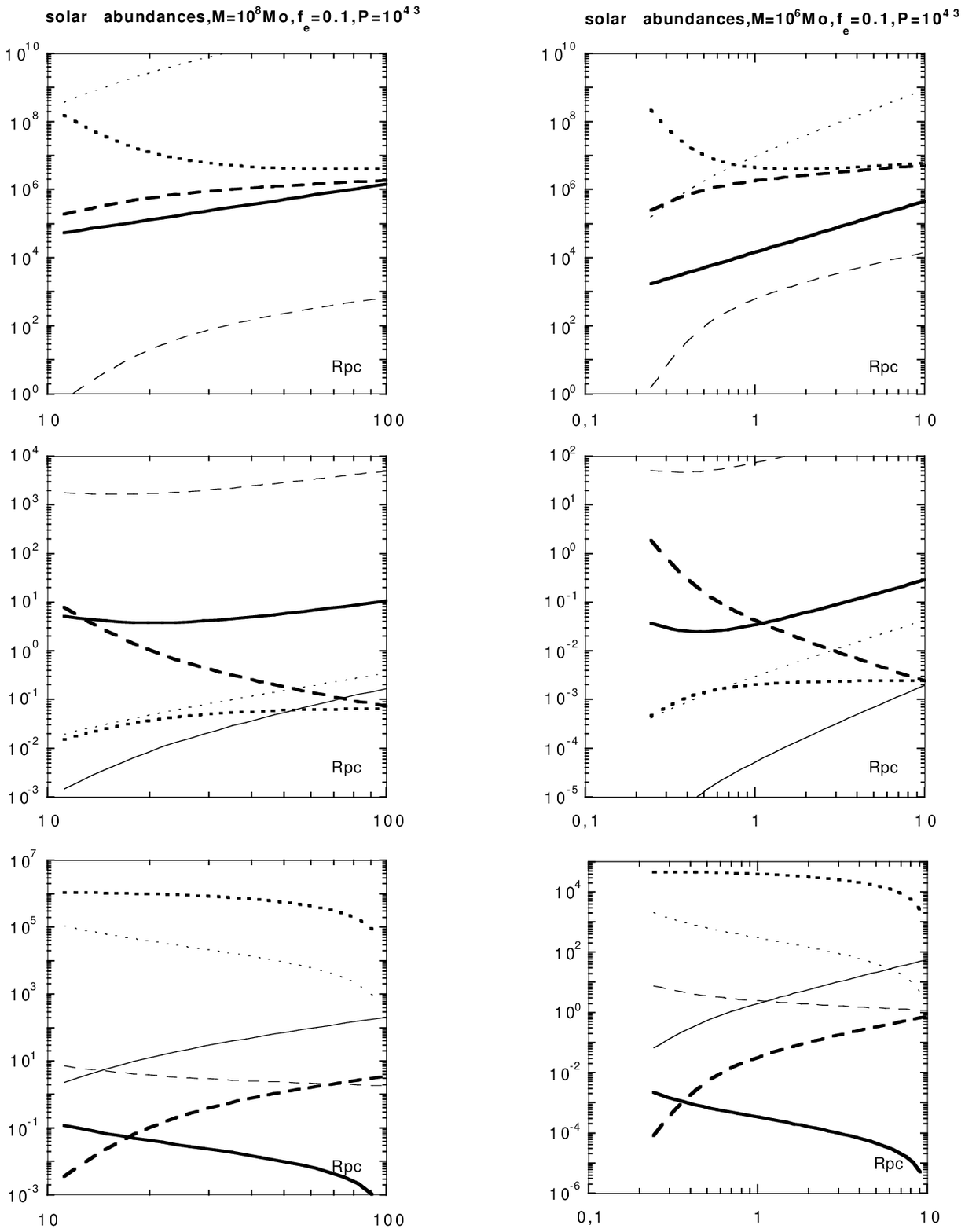}
\caption{Same as Fig.2, for other cases}
\label{fig5-gaz-etoiles}
\end{figure*}

\begin{figure*}
\psfig{figure=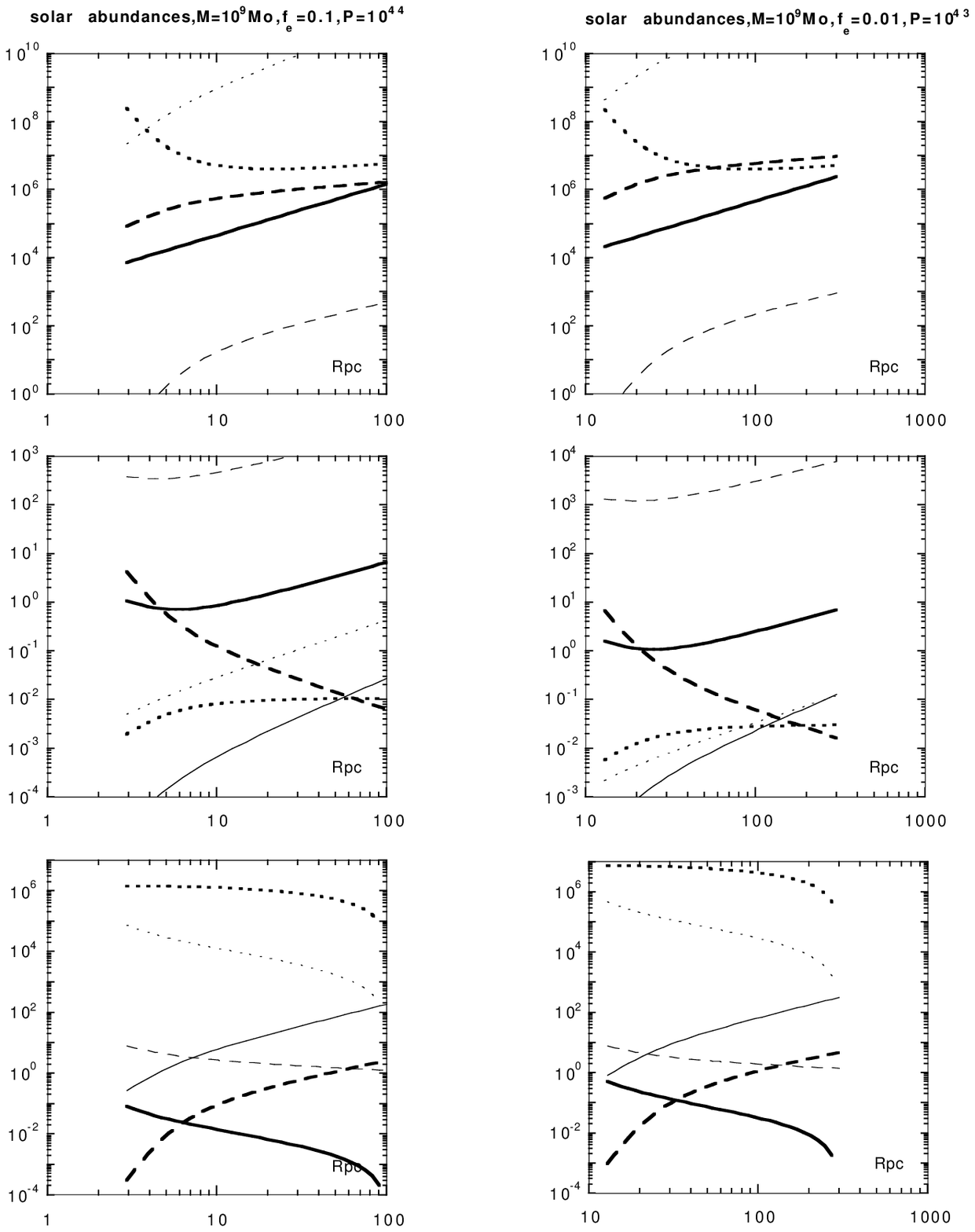}
\caption{Same as Fig.2, for other cases}
\label{fig6-gaz-etoiles}
\end{figure*}  

Figs. \ref{fig2-gaz-etoiles} to \ref{fig6-gaz-etoiles} display the 
results of a few computations. 
The luminosity of the stars is set equal to 10$^{38}$ ergs
s$^{-1}$, their mass is 30 M$_{\odot}$, and we take 3 
$10^6$ yrs for $t_{\rm lifetime}$. The total momentum given by one 
supernova is generally assumed equal to 10$^{43}$ g cm  s$^{-1}$, 
but in a few cases we use a value 10 times larger to show 
the influence of this fundamental parameter. For each  couple black 
hole mass - Eddington factor, we give two 
figures in order to check that the required conditions are 
fulfilled, and one figure displaying some interesting parameters of 
the model. 

First are displayed the characteristic times, the free fall time 
$1/\Omega$, $t_{\rm 
trans}$, $t_{\rm accr}$(10M$_{\odot}$) (we recall 
that it is an overestimation), $t_{\rm cool}$ and $t_{\rm 
migr}$(10M$_{\odot}$). The figures show that, except for small black 
hole masses and large accretion rates,  the migration time is much larger 
than the typical stellar evolution time, so migration does not prevent 
supernovae explosions. Except for large black hole masses and
accretion rates, the free fall time is smaller by about 2 orders of 
magnitude than the mass transport time, which means that 
$\sqrt{1-Q_{\rm g}^2}$ could be as small as 10$^{-2}$ (cf. Eq. 
\ref{eq-tform}). 
The assumption that the gas is marginally unstable is therefore
self-consistent.  
Finally the accretion time is at most of the order of a few 10$^{6}$ 
yrs, and we have seen that it is probably strongly overestimated.

A consequence of the large migration time is that the neutron stars 
left after the supernova explosions do not migrate towards the black 
hole. If they are not ejected from the disk (which should be the case if 
the explosions are asymmetrical or if they occur in binary stars), they 
will be able to 
reaccrete, and this phenomenon should lead to exotic massive 
stars, and probably to supernovae explosions more energetic than 
in normal galactic conditions.

We give then several parameters: $\epsilon\sqrt{1-Q_{\rm 
g}}$,  $m_{\rm gap}$, $V_{\rm rad}/V_{\rm Kep}$, the ratio of the 
rate of supernovae to the 
maximum allowed rate, the  ratio  of the number of stars to the 
maximum number allowed by the tidal effect,  and the ratio  of the 
number of stars to the 
maximum number allowed by the HII regions. These parameters
allow to check that the conditions required by the model are 
fulfilled. First we see that the minimum mass for opening a gap by 
induced density waves, $m_{\rm gap}$, is always very large, so no gap 
is opened even by large mass stars. The ratio of the radial to 
the Keplerian velocity is always small. One can also check from the 
product 
$\epsilon\sqrt{1-Q_{\rm g}}$ that a value of $\epsilon$ of the order 
of 
10$^{-3}$ is compatible with the condition $\sqrt{1-Q_{\rm g}^2}\ge 
10^{-2}$. The number of stars is always smaller than the number 
allowed by the HII regions
and by the tidal effect (except sometimes very close to the inner 
radius of the
domain). The most constraining condition comes from the maximum allowed 
number of supernovae. 
In several cases the number of supernovae is close to this maximum number, and
we show also  some cases it exceeds this number in the whole domain: $M=10^8$ M$_{\odot}$
with  $f_{\rm E}=0.1$ for primeval 
and solar abundances, $M=10^9$ M$_{\odot}$ with $f_{\rm E}=0.01$ for 
primeval abundances. To
summarize, the cases where the number of supernovae is confortably 
smaller than the maximum
number  are only those with a small
black hole mass. However we have seen that the surface covered by 
supernovae is probably
overestimated by Eq. \ref{eq-SNrate}, so we are confident that the 
cases where the rate of supernovae is slightly larger than the 
maximum allowed number are also viable.

Finally the figures show some interesting physical parameters: the scale height and the midplane
temperature of the disk,  the mass of the initial fragments $m_{\rm frag}$, the gaseous 
mass of the disk, the total number of stars, and the total supernova 
rate. The last three quantities are integrated from the external edge.

First one can easily check that the disk is geometrically thin, with 
an aspect ratio $H/R$ of the order of 10$^{-3}$ to 10$^{-2}$, and is 
always flaring (it is why we take into account external 
illumination in the heating, although it is not important). The 
midplane temperature is of the order of a few hundred degrees, 
justifying our approximations for the opacity and the cooling rate. 
The initial mass of the fragments increases strongly with the radius. 
We recall that since it depends on $H^3$ and due to our 
approximations on the opacity, it is the least correctly determined 
quantity, but this is of no consequence on the models. An interesting 
point to note is that for high black hole masses, the initial mass of 
the fragments reaches a few solar masses at the periphery of the 
domain. The total mass of the disk (actually of the domain where the 
model is valid) is much smaller than the black hole mass. The number 
of stars is small, corresponding to an amount of mass locked in stars 
of the order of that of the gas. Finally the total supernova rate, 
which is entirely dominated by the inner regions, varies from 
10$^{-3}$ to 1 M$_{\odot}$ per year. We shall come back to this 
point in the next section. 

We have not mentioned the effects of the stellar winds. 
Using Eq. \ref{eq-Rs} one finds that they are less constraining than the 
supernovae in the inner regions of the disk, and generally more constraining 
in the outer regions. We have already mentioned that in the 
case of primeval abundances, stellar winds are certainly much less powerful 
than in the case of normal abundances. More generally the energy and 
momentum of 
the winds are certainly the least known parameters in this problem, 
and we prefer not to rely too much on it.  

A first result is that the regions of interest are now
located about 10 times further from the black hole than for the pure 
gaseous disk, owing to the increased heating. A second result is that 
the model works better
when the accretion rate is smaller, and actually we have not 
found solutions for high mass black holes and $f_{\rm E}$ equal to 
unity, the reason being that the number of supernovae is then too large. 
The model works also better if the momentum provided by 
the 
supernovae is increased, since the number of stars required to transport 
the mass is smaller. Finally one must note that the model works as 
well with primeval or solar abundances.

\section{Some implications of the model}

Although the model is reminiscent of the``starburst AGN" model (cf. Terlevich and Melnick
1985 and subsequent papers), it is quite different, as the presence of the central black hole is
a fundamental ingredient. One should not forget also that though the mass accretion rate is
of the order of the outflowing mass due to supernovae, the luminous enrergy released in
supernovae is much smaller than that of the central regions of the disk. So the only
observational manifestation of the supernovae explosions would be expected in nearby low
luminosity AGN, which should display an increase of flux in the optical range every 10$^{3}$
years. Since the number of known local AGN is now of the order of a few thousands, one could
expect indeed to see a supernova exploding within a few parsecs from the center of an 
AGN, and this would be the best
check of this model.

This model has several consequences. Beside the fact that it solves 
the problem
of  the mass transport in the intermediate region of the disk, it could 
give an explanation for the
high  velocity metal enriched outflows implied by the presence of the 
Broad Absorption 
Lines (BALs) in quasars. This problem is discussed in details in 
Collin (1998), so we recall here only a few points.

There are strong observational evidences for metallicities
larger than solar (or at least solar) in the central few parsecs of quasars up to $z\ge 
4$ (see the recent systematic study of Hamann 1997). This enriched 
material is flowing out of
the central regions with a high velocity, of the order of $c$/30, as 
observed in BAL
 QSOs, which constitute about 10$\%$ of the total number of radio 
quiet
QSOs. The phenomenon is generally interpreted as an outflow existing 
in all quasars, but limited
to an opening angle $\sim 4\pi$/10. It is quite difficult to estimate 
the rate of outflowing mass,
it is between one hundredth of the accretion rate and the accretion 
rate.

Comparing the observed mass rate with the rate of supernovae, each 
releasing about 10 
M$_{\odot}$ of metals out of the nucleus, one sees that the mass 
outflow rate due to the supernovae accounts easily for the 
observations. 
Our computations show indeed that the rate of supernovae is equal to 
a few 10$^{-2}$ yr$^{-1}$
for a quasar black hole (see Fig. \ref{fig6-gaz-etoiles}), i.e. an 
outflow of metals
close to the accretion rate. The  observed velocities are also easily accounted 
for by 
the expanding shells,  and
finally the location of the phenomenon is in  agreement with 
observations. Relatively larger
enrichment in some  elements like N and Fe are observed, and could be 
explained by the oddness
of the  stars formed in a particular environment. It would actually 
be  interesting to study the
evolution of stars formed in such conditions to determine 
their final state before
the explosion.  Finally the fact that the opening angle of the BAL 
region is equal to 
a small fraction of 4$\pi$ is easily explained, the 
ejection taking place mainly in a cone in the direction of the disk axis. 
Note that in this model the
metallicity of the gas fuelling the black hole can be very small, while 
the observed outflow is always enriched.

A second  outcome of our mechanism is to account for a pregalactic 
enrichment, if massive
black holes are created early in the process of galaxy  formation, 
and if galaxy formation
takes place through an hierarchical scenario (Silk and Rees, 1998). 
Once the supernova shells are
ejected from the accretion disk, their fate  is determined by the 
mass of the host protogalaxy.
Massive galaxies will retain their gas, and the supernova shells will 
compress the interstellar 
medium, trigger star formation like in the interstellar medium, and 
induce a starburst. This is
an ``inside-outside" scenario opposite to the starburst scenario of 
Hamann and Ferland (1993),
where the central massive black hole grows by accreting already 
enriched gas during and after the
starburst, and is therefore an ``outside-inside" one. A fraction of
the enriched gas should escape from the galaxy, not only due to the 
nuclear supernovae
explosions, but also to the induced starburst.

Small galaxies will not retain their gas, which will escape with the
enriched gas produced by the supernovae. It will 
pollute the intergalactic medium (IGM). In particular, if the formation of the black holes 
precedes the formation of galaxies, it will lead to a pregalactic 
enrichment 
of the IGM. If the universe at high redshift is dominated by a 
homogeneous
population of compact and spheroidal galaxies (Steidel et al 1996) 
which 
are the progenitors of massive galaxies, the enrichment of IGM 
induced by the black holes would be quite homogeneous and the 
appeal to a population III stars would not be necessary.  

 According to our computations the mass of metals 
ejected by 
 the disk is of the order of the mass of the 
 black hole itself.
 We can therefore estimate the {\bf minimum} enrichment of IGM due to black 
holes, simply 
 taking the
integrated comoving mass density of {\bf observed} quasars, which 
corresponds 
to about 10$^{-6}$ of the closure density (Soltan, 1982, and further 
studies). If these black holes have
a typical mass of 10$^8$ M$_{\odot}$, the present mechanism will 
provide about 10$^{-6}$ of the
closure density in metals, i.e. after mixing with the IGM, an average 
metallicity of a few 10$^{-3}$
$\Omega({\rm IGM})_{0.02}Z_{\odot}$, close to the metallicity 
observed in the 
L$\alpha$ forest which constitutes the main fraction of the IGM.

Finally, our Galaxy is presently not active and the black hole in the 
center has a small accretion rate ($< 
10^{-4}$  M$_{\odot}$ yr$^{-1}$), so it has most probably accreted 
a large fraction of its mass (2 
10$^6$ M$_{\odot}$) during an 
early period. The previous estimation leads to an ejection of a few $\sim 
10^{4}$ M$_{\odot}$ of metals. After mixing with a hydrogen halo of 
10$^{11}$  M$_{\odot}$, it gives a metallicity of a few 10$^{-5}$ solar, 
close to that observed in the oldest halo stars. 

\section{Final remarks}

In this paper we have explored a new model of accretion disk, made of stars 
and gas, with a permanent formation and death
of stars, and we
have shown that stationary solutions exist for primeval or 
solar abundances  
between 0.1 and 10 pc for a black hole mass of $10^6$M$_\odot$, and between 1 and 100 pc for
a black hole mass of $10^8$M$_\odot$ or more.

 However this model 
works only for a relatively low Eddington 
factors, $\le 0.1$. This raises a severe problem, as 
some quasars can be two orders of magnitudes more luminous than those 
considered here or the black 
hole mass would have to be 
very large to account for the luminosity ($\ge 10^{10}$ M$_{\odot}$), which 
seems unlikely. Moreover during the growing 
phase of quasars, the average accretion rate should be also at least of the order 
of $\dot{M}_{\rm crit}$/10 ($\dot{M}_{\rm crit}$ being the critical 
rate), or the growth of the black hole 
would take a time larger than allowed by the existence of high 
redshift quasars, since the e-folding time for the growing of a black hole is 4 10$^7/f_{\rm 
 E}$ yrs (independently of the mass). In luminous quasars, the number of supernovae 
required to transport the angular momentum is too large, and the disk 
should become entirely paved with 
cavities surrounded by shocks. Is the disk then destroyed? Probably not, as the potential of
the black hole should still  maintain a fraction of the gas
close to  the equatorial plane. It is however possible that the thin 
disk is replaced by a thick disk (a ``torus"). In any case the transport of mass and 
angular momentum by the supernovae should be very efficient, but the 
disk 
cannot be treated like in this paper as a homogeneous gas close to 
marginal unstability. There would be some highly compressed 
regions where star formation should take place (but not 
necessarily be followed by subsequent accretion on the protostars, so 
high mass stars might not dominate the mass function as in our model). 
Clearly this case would deserve a serious study, and
numerical simulations would probably be the best way to tackle the 
problem.

Actually the most realistic scenario is that the mass inflow from the periphery of the 
disk is variable with time, as it would be the case if it is achieved by 
large molecular complexes comparable to those present near the 
Galactic center. If the mass inflow is variable with
time, there will be ``low states" and ``high states" where the disk
is  alternatively ``quiescent",  and  ``active", i.e. highly perturbed by an intense
supernova activity, and star 
formation will occur in successive ``waves" propagating from the outer 
to the inner regions. An inflow of matter will  
induce an increased gas density in an outer ring. Before the stars 
form, accrete and
evolve to supernovae, the transfer of mass will not take place, and there will 
be an accumulation of gas in the ring. After a few 10$^6$ years,
supernovae will induce mass transport towards an inner ring, while the 
outer ring will be cleared out of its gas until a new mass inflow.  Note 
that during  the ``active" phases
the mass inflow at the
periphery could be super Eddington. In 
this case our description should be modified to take into account the
non-stationarity of the process, an one would expect that the corresponding averaged
momentum transport (i.e. the accretion rate) could be much larger than in 
the stationary case, as it is not limited by a maximum allowed rate of 
supernovae.

Finally we should mention that an important improvement of the model 
would be to use better opacities, in particular to take into account 
the frequency dependence of the opacity. In this context it would also be interesting 
to study intermediate cases 
between a solar and a zero metallicity. 

\begin{acknowledgements}

We are grateful to T. Montmerle, M. Rozyccka,  M. Tagger and B. Volmer for  fruitful discussions.
 
\end{acknowledgements}

\end{document}